\documentclass[]{article}
\usepackage{axodraw}
\usepackage{jheppub}

\def\at{\alpha_t}
\def\ab{\alpha_b}
\def\as{\alpha_s}
\def\atau{\alpha_{\tau}}

\def\oatab{{\cal O}(\at\ab)}
\def\oatas{{\cal O}(\at\as)}
\def\oabas{{\cal O}(\ab\as)}

\def\oatq{{\cal O}(\at^2)}
\def\oabq{{\cal O}(\ab^2)}
\def\oatauq{{\cal O}(\atau^2)}
\def\oabatau{{\cal O}(\ab \atau)}

\def\SOFTSUSY{{\tt SOFTSUSY}}
\def\SUSPECT{{\tt SUSPECT}}

\def\ISAJET{{\tt ISASUGRA}}

\def\code#1{\small{\tt #1}\normalsize}
\def\gev{~\mbox{GeV}}
\def\half{\frac{1}{2}}
\def\hepph#1{{\tt arXiv:hep-ph/0406166}}

\title{{\SOFTSUSY}{\tt 3.3.3}: a program for calculating supersymmetric spectra}

\author[a]{B.C.~Allanach}
\affiliation[a]{DAMTP, CMS, University of Cambridge, Wilberforce road, Cambridge, CB3 0WA,
United Kingdom}
\emailAdd{B.C.Allanach@damtp.cam.ac.uk}

\abstract{\SOFTSUSY~is a program which accurately calculates the spectrum of superparticles in the CP-conserving Minimal Supersymmetric Standard Model (MSSM), with a full flavour mixing structure. The program solves the renormalisation group equations with theoretical constraints on soft supersymmetry breaking terms provided by the user. Weak-scale gauge coupling and fermion mass data (including one-loop finite MSSM corrections) are used as a boundary condition, as well as successful radiative electroweak symmetry breaking. The program can also calculate a measure of fine-tuning. The program structure has been designed to easily generalise to extensions of the MSSM\@. This article serves as a self-contained guide to prospective users, and indicates the conventions and approximations used.}

\begin{document}
\maketitle
\section{Introduction}

The Minimal Supersymmetric Standard Model (MSSM) provides an attractive
weak-scale extension to the Standard Model. As well as solving the gauge
hierarchy problem, it can be motivated by more fundamental models such as 
various string theories or supersymmetric grand unified theories. 
The MSSM provides a rich and complicated
phenomenology. It predicts many states  extra to
the Standard Model (sparticles) and
their indirect empirical 
effects and direct detection are vital for
verification of the MSSM\@.
Models that are more fundamental than
the MSSM can provide stringent constraints upon
the way supersymmetry (SUSY) is broken, with important implications for the
spectrum which in turn affects the signatures available in experiments.
It is therefore desirable to construct a calculational tool which may 
provide a spectrum and couplings of the MSSM sparticles so that studies 
of the capabilities of colliders, extraction of 
high scale parameters (if a signal is observed) and studies of constraints on
the models are enabled. We present such a tool (\SOFTSUSY) in this article.

\subsection{The Nature of the Physical Problem}

The determination of sparticle masses and couplings of SUSY particles in the
R-parity conserving MSSM is the basic problem. Low energy data on Standard
Model fermion masses,
gauge couplings and electroweak boson masses are to be used as  a constraint.
SUSY radiative corrections to these inputs from sparticle loops depend upon
the sparticle spectrum, and must be calculated.
Theoretical constraints on the SUSY breaking parameters from an underlying
theory
are often imposed at a high renormalisation scale, perhaps resulting from a
super gravity or string theory. Often, the theoretical constraints drastically
reduce the number of free parameters in the SUSY breaking sector (which
numbers over 100 in the unconstrained case). These constraints then make
phenomenological analysis tractable by reducing the dimensionality of
parameter space sufficiently so that parameter scans over a significant volume
of parameter space are possible. Finally, the MSSM parameters must also be
consistent with a minimum in the Higgs potential which leads to the observed
electroweak boson masses.

 This problem has been addressed many times before in the
literature (see for
example~\cite{Ross:1992tz,Barger:1994gh,Pierce:1997zz,Djouadi:1998di,Baer:1999sp}), 
with  varying degrees of accuracy in each
part of the calculation. It is our purpose here to provide a tool which will
solve the problem with a high accuracy, including state-of-the-art corrections.
Similar problems in the context of MSSM extensions\footnote{By MSSM extension,
we mean an extension applicable near the weak scale.} have also been
studied. In anticipation of new forms of SUSY breaking constraints and new
MSSM extensions, we designed the tool to be flexible and easily extended.

\subsection{The Program}

\SOFTSUSY~has been written in object-oriented C++ but users may use an
executable program with input either in the SUSY Les Houches
Accord~\cite{lhacc} 
format or from command-line arguments. For users wishing to call
\SOFTSUSY~from their own programs, the user interface is designed to be 
C-like to aid users that are unfamiliar with object orientation.
Accuracy and ease of generalisation have taken priority over running speed in the
design. For example,
full three family mass and Yukawa matrices may be employed, rather than
the more usual dominant third family approximation. The
publicly released codes \ISAJET~(which comprises part of the 
{\tt ISAJET}~package~\cite{Baer:1999sp}) and \SUSPECT~\cite{Djouadi:1998di}
use the dominant third family approximation, for example. 
The full three-family choice slows the renormalisation group evolution
significantly, but  
will facilitate studies of sparticle or quark mixing. The running time is not
foreseen as a bottleneck because it is a matter of a couple of seconds on a
modern PC, and will certainly be negligible compared to any Monte-Carlo
simulation of sparticle production and decay in colliders.
It is possible for the user to specify their own high scale boundary
conditions for the soft SUSY 
breaking parameters without having to change the \SOFTSUSY~code.
For the convenience of most users however, the most commonly used
high-scale boundary conditions are included in the package.

The code can be freely obtained from the \SOFTSUSY~web-page, which
currently resides at the address
\begin{quote}
\href{http://projects.hepforge.org/softsusy/}
{http://projects.hepforge.org/softsusy/} 
\end{quote}
Installation instructions and more detailed technical documentation of the
code may also be found there.

\SOFTSUSY~is
a tool whose output could be used for studies of MSSM sparticle
searches~\cite{Allanach:2000ii} by using event generators such as {\tt
  HERWIG}~\cite{Corcella:2000bw}, or other more theoretical or astrophysical
studies. 
For a review of SUSY tools on offer (which may use the output from \SOFTSUSY),
see 
Ref.~\cite{Allanach:2008zn}. 

\subsection{Aims and Layout}

The main aims of this article are to provide a manual for the use of \SOFTSUSY,
to describe the approximations employed and to detail the notation used in
order to
allow for user generalisation. There have been other articles published on the
comparison of the calculation in \SOFTSUSY~with those of other
codes~\cite{comparison}, and so we decline from including such
information here.

The rest of this paper proceeds as follows: 
the relevant MSSM parameters are presented in
sec.~\ref{sec:notation}. The approximations employed are noted in
sec.~\ref{sec:calculation}, but brevity requires that they are not
explicit. However, a reference is given so that the precise formulae utilised
may be obtained in each case. The algorithm of the calculation is also
outlined.
Technical information related to running and extending the program is
placed in appendices.
A description of how to run the command-line interface is given in
appendix~\ref{sec:run}, including information on the input-file.
Appendix~\ref{sec:prog} gives an
example of a main program, useful if the user wants to
call \SOFTSUSY~from his or her main program.
The sample output from this program is displayed and explained in
appendix~\ref{sec:output}. 
The use of switches and constants is explained
in appendix~\ref{sec:switches}. Finally, in appendix~\ref{sec:objects}, 
a description of the relevant objects and their relation to each other is
presented. 

\section{MSSM Parameters \label{sec:notation}}

In this section, we introduce the MSSM parameters in the
\SOFTSUSY~conventions. Translations to the actual variable names used in the
source code are shown in appendix~\ref{sec:objects}.

\subsection{Supersymmetric Parameters \label{susypars}}
The chiral superfields of the MSSM have the 
following $G_{SM}=SU(3)_c\times SU(2)_L\times U(1)_Y$ quantum numbers
\begin{eqnarray}
L:&(1,2,-\half),\quad {\bar E}:&(1,1,1),\qquad\, Q:\,(3,2,\frac{1}{6}),\quad
{\bar U}:\,({\bar 3},1,-\frac{2}{3}),\nonumber\\ {\bar D}:&({\bar 3},1,\frac{1}{3}),\quad
H_1:&(1,2,-\half),\quad  H_2:\,(1,2,\half).
\label{fields}
\end{eqnarray}
Then, the superpotential is written as
\begin{eqnarray}
W&=& \epsilon_{ab} \left[ (Y_E)_{ij} L_i^b
H_1^a {\bar E}_j + (Y_D)_{ij} Q_i^{bx} H_1^a {\bar D}_{jx} +
(Y_U)_{ij} Q_i^{ax} H_2^b {\bar U}_{jx}  + \mu  H_1^b H_2^a\right]
\label{superpot}
\end{eqnarray}
Throughout this section, we denote an $SU(3)$ colour index of the
fundamental representation by 
$x,y,z=1,2,3$. The $SU(2)_L$ fundamental representation indices are
denoted by $a,b,c=1,2$ and the generation indices by $i,j,k=1,2,3$.
$\epsilon_{ab}=\epsilon^{ab}$ is the totally antisymmetric tensor, with
$\epsilon_{12}=1$. 
Note that the sign of $\mu$ is identical to the one in 
\ISAJET~\cite{Baer:1999sp},
but is in the opposite convention to ref.~\cite{Pierce:1997zz}.
Presently, real Yukawa couplings only are included.
All MSSM running parameters are
in the $\overline{DR}$ scheme. The Higgs vacuum expectation values (VEVs) are
$\langle H_i^0 \rangle = v_i / \sqrt{2}$ and
$\tan\beta=v_2/v_1$. $g_i$ are the MSSM $\overline{DR}$ gauge couplings and 
$g_1$ is defined in the Grand Unified normalisation $g_1 = \sqrt{5/3} g'$,
where $g'$ is the Standard Model hypercharge gauge coupling.
Elements of fermion mass matrices are given by
\begin{equation} \label{yuksaway}
(m_u)_{ij} = \frac{1}{\sqrt{2}} (Y_U)_{ij} v_2, \qquad 
(m_{d,e})_{ij} = \frac{1}{\sqrt{2}} (Y_{D,E})_{ij} v_1
\end{equation}
for the up quark, down quark and charged lepton matrices respectively.

\subsection{SUSY Breaking Parameters \label{sec:susybreak}}
We now tabulate the notation of the soft SUSY breaking parameters. The
trilinear scalar interaction potential is
\begin{equation}
V_3 = \epsilon_{ab}
\left[
\tilde{Q}_{i_L}^{xa} (U_A)_{ij}  \tilde{u}_{jx_R} H_2^b + 
\tilde{Q}_{i_L}^{xb} (D_A)_{ij}  \tilde{d}_{jx_R} H_1^a + 
\tilde{L}_{i_L}^{b} (E_A)_{ij}  \tilde{e}_{j_R} H_1^a + H.c.\right],
\end{equation}
where fields with a tilde are the scalar components of the superfield with the
identical capital letter. 
Also,
\begin{equation}
(A_{U,D,E})_{ij} = (U_A,D_A,E_A)_{ij} / (Y_{U,D,E})_{ij} 
\end{equation}
(no summation on $i,j$) are often referred to in the literature as soft
$A$-parameters.

The scalar bilinear SUSY breaking terms are contained in the potential
\begin{eqnarray}
V_2 &=& m_{H_1}^2 {{H_1}_a}^* {H_1^a} + m_{H_2}^2 {{H_2}_a}^* {H_2^a} +
{\tilde{Q}^*}_{ixa} (m_{\tilde Q}^2)_{ij} \tilde{Q}_j^{xa} +
{\tilde{L}^*}_{ia} (m_{\tilde L}^2)_{ij} \tilde{L}_j^{a}  + \nonumber \\ &&
 \tilde{u}_i^{x} (m_{\tilde u}^2)_{ij}  {\tilde{u}^*}_{jx} +
\tilde{d}_i^{x} (m_{\tilde d}^2)_{ij}  {\tilde{d}^*}_{jx} +
\tilde{e}_i (m_{\tilde e}^2)_{ij} {\tilde{e}^*}_{j} +
\epsilon_{ab} (m_3^2 H_2^a H_1^b + H.c.).
\end{eqnarray}
For a comparison of these conventions with other popular ones in the
literature, see Table~\ref{tab:conv}. In the table, we compare the
\SOFTSUSY~conventions with the 
SUSY Les Houches Accord~\cite{lhacc} and ref.~\cite{mandv} (Martin and Vaughn).
\begin{table}\begin{center}
\begin{tabular}{ccc}
\label{tab:conv}
\SOFTSUSY & SLHA & Martin and Vaughn \\ \hline
$Y^{U}$ & $Y^{U}$ & $({Y^{U}})^T$ \\
$Y^{D,E}$ & $Y^{D,E}$ & $(-{Y^{D,E}})^T$ \\
$U_A$ & $T_{U}$ & $h_{U}^T$ \\
$D_A, E_A$ & $T_{D,E}$ & $(-h_{D,E})^T$ \\
$m_{{\tilde Q},{\tilde L}}^2$ & $m_{{\tilde Q},{\tilde L}}^2$ & $m_{{\tilde
    Q},{\tilde L}}^2$ \\
$m_{{\tilde u},{\tilde d},{\tilde e}}^2$ & ${m_{{\tilde u},{\tilde d},{\tilde
    e}}^2}^T$ &
$m_{{\tilde u},{\tilde d},{\tilde e}}^2$\\
$\mu$ & $\mu$ & $\mu$ \\
$m_3^2$ & $B \mu$ & $B$  \\
$M_i$ & $M_i$ & $M_i$ \\
$m_{H_{1,2}}^2$ & $m_{H_{1,2}}^2$& $m_{H_{d,u}}^2$ \\
\end{tabular}
\caption{Comparison of conventions between \SOFTSUSY~and the literature. Note
  that simple models of SUSY breaking, for example the CMSSM, will have
  negative gaugino masses $M_i$ in the \SOFTSUSY~conventions.}
\end{center}\end{table}

Writing the bino as ${\tilde b}$, ${\tilde w}^{A=1,2,3}$ as the
unbroken-SU(2)$_L$ 
gauginos and ${\tilde 
g}^{X=1\ldots8}$ as the gluinos, the gaugino mass terms are contained in the
Lagrangian 
\begin{equation}
{\mathcal L}_G = \frac{1}{2} \left( M_1 {\tilde b}{\tilde b} + M_2 {\tilde w}^A{\tilde w}^A
+ M_3 {\tilde g}^X {\tilde g}^X \right) + h.c.
\end{equation}

\subsection{Tree-Level Masses \label{sec:tree}}
Here we suppress any gauge indices and follow the notation of
ref.~\cite{Pierce:1997zz} closely.
The Lagrangian contains the neutralino mass matrix as
$-\frac{1}{2}
{\tilde\psi^0}{}^T{\cal M}_{\tilde\psi^0}\tilde\psi^0$ + h.c., where
$\tilde\psi^0 =$ $(-i\tilde b,$ $-i\tilde w^3,$ $\tilde h_1,$ $\tilde
h_2)^T$ and
\begin{equation}
{\cal M}_{\tilde\psi^0} \ =\ \left(\begin{array}{cccc} M_1 & 0 &
-M_Zc_\beta s_W & M_Zs_\beta s_W \\ 0 & M_2 & M_Zc_\beta c_W &
-M_Zs_\beta c_W \\ -M_Zc_\beta s_W & M_Zc_\beta c_W & 0 & -\mu \\
M_Zs_\beta s_W & -M_Zs_\beta c_W & -\mu & 0
\end{array} \right). \label{mchi0}
\end{equation}
We use $s$ and $c$ for sine and cosine, so that
$s_\beta\equiv\sin\beta,\ c_{\beta}\equiv\cos\beta$ and $s_W (c_W)$ is
the sine (cosine) of the weak mixing angle.  
The 4 by 4 neutralino mixing matrix is an orthogonal matrix $O$ with real
entries, 
such that $O^T {\cal M}_{\tilde\psi^0} O$ is diagonal.
The neutralinos $\chi^0_i$ are defined such that their absolute masses
increase with increasing $i$.
Some of their mass values can be negative. 

We make the identification
${\tilde w}^\pm = ({\tilde w^1} \mp i{\tilde w^2} ) / \sqrt{2}$ for the charged
winos and ${\tilde h_1^-}, {\tilde h_2^+}$ for the charged higgsinos.
The Lagrangian contains the chargino mass matrix as
$-{\tilde\psi^-}{}^T{\cal M}_{\tilde\psi^+}\tilde\psi^+ + h.c.$,
where~$\tilde\psi^+ = (-i\tilde w^+,\ \tilde h_2^+)^T,\ \tilde\psi^-=
(-i\tilde w^-,\ \tilde h_1^-)^T$ and
\begin{equation}
{\cal M}_{\tilde\psi^+}\ =\ \left( \begin{array}{cc} M_2 &
\sqrt2\,M_Ws_\beta\\\sqrt2\,M_Wc_\beta & \mu\end{array}\right).
\label{mchi}
\end{equation}
This matrix is then diagonalised by 2 dimensional rotations through angles
$\theta_L, \theta_R$ in the following manner:
\begin{equation}
\left( \begin{array}{cc} 
c_{\theta_L} & s_{\theta_L} \\
-s_{\theta_L} & c_{\theta_L} \\
\end{array} \right)
{\cal M}_{\tilde\psi^+}
\left( \begin{array}{cc} 
c_{\theta_R} & -s_{\theta_R} \\
s_{\theta_R} & c_{\theta_R} \\
\end{array} \right) =
\left( \begin{array}{cc}
m_{\chi_1}^+ & 0 \\
0 & m_{\chi_2}^+
\end{array}
\right)
\end{equation}
where $m_{\chi_i}^+$ could be negative,
with the mass parameter of the lightest chargino being in the top left hand
corner. 

At tree level the gluino mass, $m_{\tilde g},$ is given by $M_3$.

Strong upper bounds upon the inter-generational scalar mixing
exist~\cite{Gabbiani:1996hi} and in the following we assume that such mixings
are negligible.
The tree-level squark and slepton mass squared values for the family $i$
are found by diagonalising the following mass
matrices ${\mathcal M}_{\tilde f}^2$ defined in the $({\tilde f}_{iL}, {\tilde
f}_{iR})^T$ basis: 
\begin{equation}
\label{sqmu} \left(\begin{array}{cc}
(m_{\tilde Q}^2)_{ii} + {m_u^2}_i + (\half - \frac{2}{3} s_W^2)M_Z^2c_{2\beta} &
m_{u_i}\left((A_U)_{ii}-\mu\cot\beta\right)\\
m_{u_i}\left((A_U)_{ii}-\mu\cot\beta\right) & (m^2_{\tilde u})_{ii} +
m_{u_i}^2 +
\frac{2}{3} s_W^2 M_Z^2c_{2\beta}
\end{array}\right)\ ,
\end{equation}
\begin{equation}
\label{sqmd}  \left(\begin{array}{cc}
(m_{\tilde Q}^2)_{ii} + m_{d_i}^2 - (\half - \frac{1}{3} s_W^2)M_Z^2c_{2\beta} &
m_{d_i}\left((A_D)_{ii}-\mu\tan\beta\right)\\
m_{d_i}\left((A_D)_{ii}-\mu\tan\beta\right) & (m^2_{\tilde d})_{ii} +
m_{d_i}^2 -\frac{1}{3} s_W^2 M_Z^2c_{2\beta}
\end{array}\right)~,
\end{equation}
\begin{equation}
\label{sqme}  \left(\begin{array}{cc}
(m_{\tilde L}^2)_{ii} + m_{e_i}^2 - (\half - s_W^2) M_Z^2 c_{2\beta} &
m_{e_i}\left((A_E)_{ii}-\mu\tan\beta\right)\\
m_{e_i}\left((A_E)_{ii}-\mu\tan\beta\right) & (m^2_{\tilde e})_{ii} +
m_{e_i}^2 - s_W^2 M_Z^2c_{2\beta}
\end{array}\right)~,
\end{equation}
$m_f, e_f$ are the mass and electric charge of fermion $f$ respectively.
The mixing of the first two families is suppressed by a small fermion mass,
which we approximate to zero.
The sfermion mass eigenstates are given by
\begin{equation}
\left(\begin{array}{cc} m_{\tilde f_1} & 0 \\ 0 & m_{\tilde f_2} 
\end{array} \right) =
\left(\begin{array}{cc} c_f & s_f\\-s_f & c_f \end{array}\right)
{\mathcal M}_{\tilde f} ^2
\left(\begin{array}{cc} c_f & -s_f\\s_f & c_f \end{array}\right)
\end{equation}
where $c_f$ is the cosine of the sfermion mixing angle,
$\cos\theta_f$, and $s_f$ the sine. 
$\theta_f$ are set in the convention that the two mass eigenstates are in no
particular order and $\theta_f \in [-\pi/4, \pi/4]$. The sneutrinos of one
family are not 
mixed and their masses are given by 
\begin{equation}
m_{\tilde \nu_i}^2 = (m_{\tilde L}^2)_{ii} + \half M_Z^2 c_{2\beta}.
\end{equation}
The CP-even gauge eigenstates $(H_1^0,\, H_2^0)$ are rotated by the angle
$\alpha$ into the mass eigenstates $(H^0\, h^0)$ as follows,
\begin{equation}
\left(\begin{array}{c}H^0\\h^0\end{array}\right) \ =
\ \left(\begin{array}{cc} c_\alpha & s_\alpha\\-s_\alpha & c_\alpha
\end{array}\right)\left(\begin{array}{c} \frac{1}{\sqrt{2}} \Re  H_1^0
  \\\frac{1}{\sqrt{2}} \Re H_2^0\end{array}\right)~.
\label{rotateh}
\end{equation}
$m_{h^0} < m_{H^0}$ by
definition, and $\alpha \in [-\pi/4, 3 \pi/4]$.
The CP-odd and charged Higgs masses are
\begin{equation}
m_{A^0}^2 = m_3^2 (\tan \beta + \cot \beta),\qquad
m_{H^\pm}^2 = m_{A^0}^2 + M_W^2
\end{equation}
at tree level.

\section{Calculation \label{sec:calculation}}

We now show the algorithm used to perform the calculation.
Standard Model parameters (fermion and gauge boson masses,
the fine structure constant $\alpha (M_Z)$, the Fermi constant from muon decay
$G_F^\mu$ and $\alpha_3(M_Z)$) are used as constraints. 
The soft SUSY breaking parameters and the superpotential parameter $\mu$ 
are then the free parameters. However, in what follows, $|\mu|$ is constrained
by $M_Z$ and $\tan \beta$ is traded for $m_3$ as an input parameter.
Therefore, the total list of unconstrained input parameters is: any
fundamental soft 
SUSY breaking parameters (except $m_3^2$), $\tan \beta$ and the sign
of $\mu$. 
First we describe the evolution of the low-energy Standard Model input
parameters below $M_Z$, then detail the rest of the algorithm.

\subsection{Below $M_Z$}

$\alpha(M_Z)$, $\alpha_s(M_Z)$ are first evolved to 1 GeV using 3 loop QCD and
1 loop QED~\cite{Gorishnii:1990zu,Tarasov:1980au,Gorishnii:1984zi} with
step-function decoupling of fermions at their running masses.
We have checked that the contribution from 2-loop
matching~\cite{Chetyrkin:1997sg} is negligible; the effect of 3-loop terms in
the renormalisation group equations
is an order of magnitude larger.
Then, the two gauge couplings and all Standard Model fermion masses except
the top mass are run to $M_Z$. The $\beta$ functions of fermion masses are
taken to be zero at renormalisation scales below their running masses.
The parameters at $M_Z$ are used as the low energy boundary condition in the
rest of the evolution.

\subsection{Initial Estimate}

The algorithm proceeds via the iterative method, and therefore an approximate
initial guess of MSSM parameters is required. 
For this, the third family $\overline{DR}$ Yukawa couplings are approximated
by 
\begin{equation}
h_t(Q) = \frac{m_t(Q) \sqrt{2}}{v \sin \beta}, \qquad
h_{b,\tau}(Q) = \frac{m_{b,\tau}(Q) \sqrt{2}}{v \cos \beta},
\end{equation}
where $v=246.22$ GeV is the Standard Model Higgs VEV and
$Q=m_t(m_t)$ is the renormalisation scale.
The $\overline{MS}$ values of fermion masses are used for this initial
estimate. 
The fermion masses and $\alpha_s$
at the top mass are obtained by evolving the previously
obtained fermion masses and gauge couplings from $M_Z$ to $m_t$ (with the same
accuracy). 
The electroweak gauge couplings are estimated by $\alpha_1(M_Z)
= 5 \alpha(M_Z) / (3 c_W^2)$, $\alpha_2(M_Z) = \alpha(M_Z) / s_W^2$. Here,
$s_W$ is taken to be the on-shell value. These two gauge couplings are then
evolved 
to $m_t$ with 1-loop Standard Model $\beta$ functions, including the effect of
a light higgs (without decoupling it). In this initial guess, no SUSY
threshold effects are calculated. The gauge and Yukawa couplings are then
evolved to the unification scale $M_X$ with the one-loop MSSM $\beta$
functions, where the user-supplied boundary 
condition on the soft terms is applied. 
Also, $\mu(M_X) =
\mbox{sgn}(\mu)\times1\gev$ and $m_3(M_X)=0$ are imposed. 
These initial values are irrelevant; they are overwritten on the next
iteration by more realistic boundary conditions.
$\mu(M_X)$ is set to be of the correct sign
because its sign does not change through renormalisation. 

The whole system of
MSSM soft parameters and SUSY couplings is then evolved to 1-loop order to
$M_Z$. 
At $M_Z$, the tree-level electroweak symmetry breaking (EWSB) conditions 
are applied~\cite{Allanach:2000ii} to predict $\mu$ and $m_3$.
The masses and mixings of MSSM super particles are then calculated at tree-level
order by using the SUSY  parameters (and $m_3$) calculated at $M_Z$.
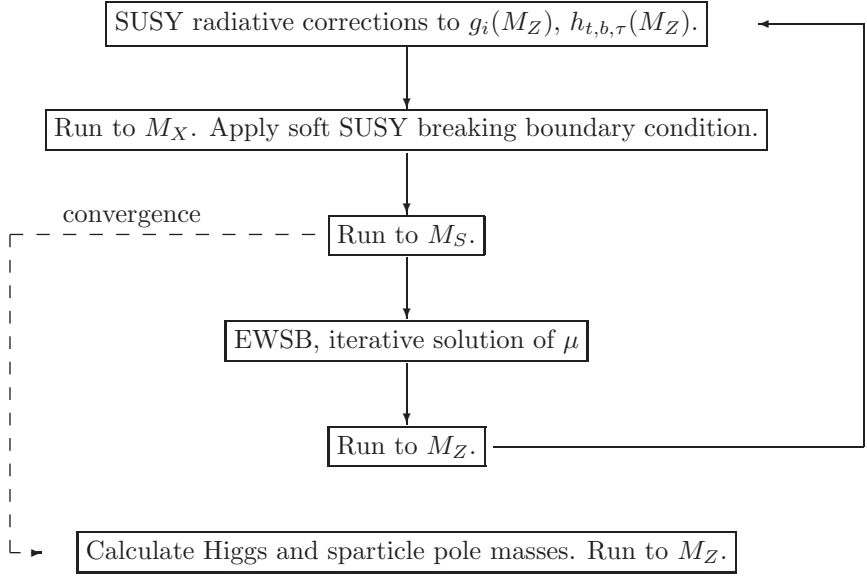
\begin{figure}\begin{center}
\label{fig:algorithm}
\begin{picture}(323,210)
\put(10,0){\makebox(280,10)[c]{\fbox{Calculate Higgs and
      sparticle pole masses. Run to $M_Z$.}}}
\put(10,40){\makebox(280,10)[c]{\fbox{Run to $M_Z$.}}}
\put(150,76.5){\vector(0,-1){23}}
\put(10,160){\makebox(280,10)[c]{\fbox{Run to $M_X$. Apply soft SUSY breaking
boundary condition.}}}
\put(150,116.5){\vector(0,-1){23}}
\put(10,80){\makebox(280,10)[c]{\fbox{EWSB, iterative solution of $\mu$}}}
\put(150,156){\vector(0,-1){23}}
\put(10,120){\makebox(280,10)[c]{\fbox{Run to $M_S$.}}}
\DashLine(115,125)(0,125){5}
\DashLine(0,125)(0,5){5}
\DashLine(0,5)(12,5){5}
\put(10,5){\vector(1,0){2}}
\put(20,130){convergence}
\put(150,196){\vector(0,-1){23}}
\put(10,200){\makebox(280,10)[c]{\fbox{SUSY radiative corrections to
$g_i(M_Z)$, $h_{t,b,\tau}(M_Z)$.}}}
\put(183,45){\line(1,0){140}}
\put(323,45){\line(0,1){160}}
\put(323,205){\vector(-1,0){40}}
\end{picture}
\caption{Iterative algorithm used to calculate the SUSY spectrum. Each step
(represented by a box) is detailed in the text. The initial step is the
uppermost one. $M_S$ is the scale at which the EWSB
conditions 
are imposed, as discussed in the text. $M_X$ is the scale at which the high
energy SUSY breaking boundary conditions are imposed.}\end{center}\end{figure}
The resulting set of MSSM parameters is then used as the initial guess for the
iterative procedure described below.

\subsection{Gauge and Yukawa Couplings}

Figure~\ref{fig:algorithm} shows the iterative procedure, starting from the the
top. The whole calculation is currently performed in the real full three family
approximation, i.e.\ all Yukawa couplings are set to be real, but quark mixing
is incorporated.
First of all, the one-loop radiative corrections are applied to the gauge
and third-family Yukawa couplings. For these, we rely heavily on
ref.~\cite{Pierce:1997zz}
by Bagger, Matchev, Pierce and Zhang (BMPZ)\footnote{Whenever a reference to
an equation in BMPZ is made, it is understood that the sign of $\mu$ must be
reversed.}.
In the threshold corrections, we use running $\overline{DR}$
masses and parameters at the relevant scale, unless denoted otherwise in the
text. 
$m_t(M_Z)$ is calculated with 2-loop QCD~\cite{avdeev} and
the full one-loop supersymmetric
contributions to $m_t(M_Z)$, including logarithmic and finite contributions
(eqs.~(D.16)-(D.18) of BMPZ):
\begin{eqnarray}
m_t(M_Z)^{\overline{DR}}_{MSSM} &=& \Sigma_t^{BMPZ} + m_t^{pole} \left( 1- \frac{\alpha_s(M_Z)}{3
  \pi} (5-3L)
+ \right. \nonumber \\ 
&& \left.
\alpha_s^2(M_Z) ( -0.538 + \frac{43}{24 \pi^2} L - \frac{3}{8 \pi^2} L^2) 
\right),
\end{eqnarray}
where $L\equiv \ln ({m_t^2(M_Z)^{\overline{DR}}_{MSSM}} / M_Z^2)$.
We denote the BPMZ corrections {\em without}\/ the one-loop QCD part as
$\Sigma_t^{BMPZ}$. 
These corrections are necessary because the region of valid EWSB is very
sensitive to $m_t$~\cite{Allanach:2000ii}. 
To calculate $m_b(M_Z)^{\overline{DR}}$, we first calculate the Standard Model
$\overline{DR}$ value from the $\overline{MS}$ one~\cite{avdeev,bottomMass}
\begin{equation}
m_b(M_Z)^{\overline{DR}}_{SM} =
m_b(M_Z)^{\overline{MS}}_{SM} \left(1 - \frac{\alpha_s^{\overline{DR}}}{3 \pi}
- \frac{23 
  {\alpha_s^{\overline{DR}}}^2}{72 \pi^2} + 
\frac{3 g^2}{128 \pi^2} + \frac{13 g_1^2}{1152 \pi^2} \right).
\end{equation}
We then add the leading one-loop supersymmetric corrections
\begin{equation}
m_b(M_Z)^{\overline{DR}}_{MSSM}=m_b(M_Z)^{\overline{DR}}_{SM} / (1 +
  \Delta_{SUSY}^b).
\end{equation}
The contributions to $\Delta_{SUSY}^b$ are
included in full from eq.~D.18 of BMPZ (neglecting the term proportional to
$e$, since that is already included in the QED calculation of the SM 
$m_b(M_Z)$).
Both finite and leading logarithmic
corrections are included.
After the Standard Model $\overline{MS}$ bar of $m_\tau$ is converted to the
$\overline{DR}$ value via
\begin{equation}
m_\tau(M_Z)^{\overline{DR}}_{SM} = m_\tau(M_Z)^{\overline{MS}}_{SM}
\left( 1 - \frac{3}{128 \pi^2} (g_1^2 - g_2^2)\right).
\end{equation}
The full one-loop MSSM corrections from the appendix 
of BPMZ 
(aside from the photon
contribution, since that has already been included in
$m_\tau(M_Z)^{\overline{MS}}_{SM}$) 
are then used to correct
$m_\tau(M_Z)^{\overline{DR}}_{SM}$:
\begin{equation}
m_\tau(M_Z)^{\overline{DR}}_{MSSM} = m_\tau(M_Z)^{\overline{DR}}_{SM}
 (1 + \Sigma_\tau).
\end{equation}
The one-loop $\overline{DR}$ values for $m_t(M_Z)$, $m_b(M_Z)$,
$m_{\tau}(M_Z)$ are then substituted with the $\overline{DR}$ value
of $v(Q)$ into eq.~(\ref{yuksaway}) to calculate the third family
$\overline{DR}$ Yukawa couplings at $M_Z$.
$v(Q)$ is run to two-loops.
The other diagonal elements of the Yukawa matrices are set by
eq.~(\ref{yuksaway}) but with fermion masses replaced by the $\overline{MS}$
values. 

The default option is to perform the calculation in the 
dominant third-family approximation, where all elements of Yukawa matrices
expect for the (3,3) elements are set to zero.
There are also options described in
appendix~\ref{sec:switches} for performing the calculation in the unmixed
3-family approximation or the fully-mixed 3-family case.
If a flavour-mixing option is chosen (see section~\ref{sec:flavour}),
the Yukawa couplings are then mixed using the ``standard parameterisation''
of the CKM matrix~\cite{Groom:2000in} with CP-violating phase set either to
zero or $\pi$, whichever results in a positive entry for $(V_{CKM})_{13}$
(also known as $V_{ub}$):
\begin{equation}
V_{CKM} = \left( \begin{array} {ccc}
c_{12} c_{13}   & s_{12} c_{13}   & p s_{13} \\
-s_{12} c_{23} -p c_{12} s_{23} s_{13} & c_{12} c_{23} -p s_{12} s_{23} s_{13} & 
s_{23} c_{13} \\
s_{12} s_{23} - p c_{12} c_{23} s_{13} & -c_{12} s_{23} - p s_{12} c_{23} s_{13} &
c_{23} c_{13} \\
\end{array} \right), \label{ckmPar}
\end{equation}
where $s_{ij}\equiv \sin \theta_{ij}$, $c_{ij}\equiv \cos \theta_{ij}$ and
$p=\pm 1$. Sign conventions are automatically chosen such that diagonal entries and entries
above the diagonal in $V_{CKM}$ are positive. Note that $V_{CKM}$ is a member
of $O(3)$, i.e.\ $V_{CKM}^{-1}= V_{CKM}^T$.
While complex phase effects are obviously not taken into account in
eq.~\ref{ckmPar}, it is hoped that the magnitudes of the main quark mixing
effects will be. In fact, using the central values for $|(V_{CKM})_{ij}|$
given by 
the particle data group~\cite{Groom:2000in}, the magnitudes of all elements in
$V_{CKM}$ in the 
first row and last column may be exactly reproduced. Of the other entries, 
$|V_{cs}|$ is accurate at the 2$\times 10^{-5}$ level and
$|V_{cd}|$ at the 4$\times 10^{-4}$ level: surely negligible for most practical
purposes. $|V_{ts}|$ is also quite accurate (to 1.2$\%$), but 
it should be noted that $|V_{td}|$ is wrong by around 50$\%$. Any
flavour physics effects sensitive to $|V_{td}|$ is therefore subject to
this large uncertainty on its value issuing from {\tt SOFTSUSY}\footnote{It is
  hoped 
in the future to include complex phases in Yukawa matrices,
sfermion soft mass squared terms and trilinear scalar couplings.}.
The up (by default), or down Yukawa couplings at $M_Z$ are mixed in the
weak eigenbasis via
\begin{equation}
(Y_U)'=V_{CKM}^T (Y^U) V_{CKM}, \qquad 
(Y_D)'=V_{CKM} (Y^D) V_{CKM}^T \label{ckm}
\end{equation}
where the primed Yukawa matrix is in the weak eigenbasis and the unprimed is
in the mass eigenbasis. 

Full one-loop corrections to $g_i(M_Z)$ are
included. 
The treatment of electroweak gauge couplings follows
from appendix C of BMPZ, and includes: two-loop corrections 
from the top, electroweak boson and the lightest CP-even Higgs. The pole value
of $m_t$ is used in the calculation of the $W$ and $Z$ self energy in order to
calculate $\sin \theta_w$, since this is what is assumed in the two-loop
corrections. 
We use the fine structure constant $\alpha(M_Z)^{\overline{MS}}$, the Z-boson
mass 
$M_Z$ and the Fermi decay constant $G_\mu$ as inputs. $M_W$ is predicted from
these inputs.
Because the EWSB constraints tend to 
depend sensitively upon $g_{1,2}(M_Z)$, accurate values for them are
determined iteratively. 
An estimate of the $\overline{DR}$ value of $s_W^2$ is used to yield a
better estimate until the required accuracy is reached (usually within 3 or
4 iterations).
The QCD coupling is input as $\alpha_s(M_Z)^{\overline{MS}}$ and
is modified by gluino, squark and top loops as in
eqs.~(2),(3) of BMPZ in order to obtain the MSSM $\overline{DR}$ value. 

\subsection{MSSM Renormalisation}

All soft breaking and SUSY parameters are then evolved to the scale
\begin{equation}
M_S \equiv \sqrt{m_{{\tilde t}_1}(M_S) m_{{\tilde t}_2}(M_S)}, \label{msusy}
\end{equation}
where~\cite{Casas:1998vh} the scale dependence of the electroweak
breaking conditions is smallest. 
Throughout the iteration described here, the renormalisation group evolution
(RGE) employs three family,
2-loop MSSM $\beta$ functions for the supersymmetric
parameters~\cite{Barger:1994gh}.
$\tan \beta$ and the Higgs VEV
parameter $v$ are also run to two-loop order in the Feynman gauge, although
the Higgs VEV RGE is missing terms ${\mathcal O}(g_2^4, g_2^2 g_1^2,
g_1^4) / (16 \pi^2)^2$~\cite{Martin:2001vx,Yamada:2001ck}. 
There is no
step-function decoupling of sparticles: this is taken into account at leading
logarithmic order in the radiative corrections previously calculated at $M_Z$
and in the calculation of the physical sparticle spectrum at $M_S$, described
below. All $\beta$ functions are real and include 3 family (and mixing)
contributions. If no flavour mixing is present in the model specified by the
user, the 2-loop parts of the RGEs switch to the dominant 3rd-family version,
where the lighter two families' Yukawa couplings are neglected.

\subsection{Electroweak Symmetry Breaking}

The Higgs VEV parameter $v(M_S)$ is set by:
\begin{equation}
v^2(M_S) = 4 \frac{M_Z^2 + \Re\Pi_{ZZ}^T(M_S)}{g_2^2(M_S) + 3 g_1^2(M_S) / 5},
\end{equation}
where
$M_Z$ is the pole $Z$ mass and
$\Pi_{ZZ}^T$ is the transverse $Z$ self-energy. 
The full one-loop EWSB
conditions at this scale are then employed to\footnote{Note that there is also
  an option to extract $m_{H_1}(M_{SUSY})$ and $m_{H_2}(M_{SUSY})$ from input
  $\mu(M_{SUSY})$ and  
$m_A(\mbox{pole})$ values, see section~\protect\ref{sec:altewsb}.}
calculate $m_3(M_S)$ and $\mu(M_S)$. $\mu(M_S)$ requires an iterative solution
because the tadpoles depend upon the value of $\mu$ assumed.
The symmetry breaking condition for $\mu$ can be phrased
as~\cite{Pierce:1997zz}
\begin{equation}
\mu^2 = \half \left( \tan 2\beta \left[m_{\bar{H}_2}^2\tan \beta
- m_{\bar{H}_1}^2  \cot \beta \right] -  M_{\bar Z}^2 \right),
\label{mucond} 
\end{equation}
where $m_{\bar{H}_i}^2 = m_{H_i}^2 - t_i/v_i$, $M_{\bar Z}^2 = M_Z^2 +
\Re\Pi_{ZZ}^T(M_Z^2)$ is the running $Z$ mass and
$t_i$ are the tadpole contributions.
The value of $\mu$ coming from the tree-level EWSB
condition (eq.~\ref{mucond}, with $\Re\Pi_{ZZ}^T=t_i=0$)
is utilised as an initial guess, then the one-loop contributions in the
tadpoles and self-energy terms are added to
provide a new value of $\mu(M_S)$. Two-loop terms 
$\oatq$, $\oabatau$, $\oabq$, $\oabas$, $\oatas$, $\oatauq$ and
$\oatab$
 are also included
in the 
tadpoles~\cite{Dedes:2002dy,Dedes:2003km}. 
The tadpole corrections are then calculated
using the new value of $\mu(M_S)$ and the procedure is repeated until it
converges to a given accuracy. $m_3(M_S)$ is then determined by input the value
of $\mu(M_S)$ into the EWSB condition
\begin{equation}
m_3^2=\frac{s_{2\beta}}{2} \left( m_{\bar{H}_1}^2 + m_{\bar{H}_2}^2 + 2 \mu^2
\right). \label{Bcond}
\end{equation}
The ensemble of MSSM parameters are then evolved using the $\beta$ functions
described above to the user supplied scale $M_X$. 
If gauge-unification has been specified as a boundary condition, the current
estimate of $M_X$ is revised to leading log order 
to provide a more accurate value upon
the next iteration:
\begin{equation}
M_X^{new} = M_X \exp \left({\frac{g_2(M_X) - g_1(M_X)}{g_1'(M_X) - g_2'(M_X)}}\right),
\label{mguteq}
\end{equation}
where primes denote derivatives calculated to 2-loop order.
The user-supplied boundary
conditions are then imposed upon the soft terms before the model is evolved
back down to $M_S$. The super particle mass spectrum is determined at this
scale. Because $\mu$ and $m_3$ are 
more scale independent at $M_S$ as opposed to some other scale, the Higgs,
neutralino and chargino masses also ought to be more scale independent 
by determining them at this scale. 

\subsection{MSSM Spectrum}

In the following description of the approximations involved in the calculation
of the super particle spectrum, it is implicit that where masses
appear, their $\overline{DR}$ values 
are employed. The running value of
$s_W(\mu)=e(\mu) / g_2(\mu)$ is also employed.
In loop corrections to sparticle masses, the Yukawa couplings of the first two
families are set to zero, being highly suppressed compared to those of the
third family. 
All sparticle masses are calculated with the full SUSY one-loop BPMZ
corrections at the scale $M_{SUSY}$. Most sparticle masses are
calculated at external momenta equal to their $\overline{DR}$ mass
$m(M_{SUSY})$. 

The physical gluino mass is calculated to full one-loop order as follows.
The running parameters are evaluated at renormalisation scale 
$\mu=M_{SUSY}$ and external momentum 
$p=M_3(\mu)$ in the following corrections:
\begin{eqnarray}
\Delta_{\tilde g} (\mu) = \frac{g_3(\mu)^2}{16 \pi^2} &&
\left( 15 + 9 \ln \left( \frac{\mu^2}{p^2} \right) - \sum_q \sum_{i=1}^2
B_1(p, m_q, m_{\tilde q_i}, \mu) + \right. \nonumber \\
&& \left. 
\sum_{q=t,b} \frac{m_q}{M_3(\mu)} s_{2 \theta_q} \left[ B_0(p, m_q, m_{\tilde
q_1}, \mu) - 
B_0(p, m_q, m_{\tilde q_2}, \mu) \right] 
\right).
\end{eqnarray}
The Passarino-Veltman functions $B_{0,1}$ are given in appendix B of BMPZ\@.
The physical gluino mass is then given by
\begin{equation}
m_{\tilde g} = M_3(M_{SUSY}) \left(1 + \Delta_{\tilde g} (M_{SUSY})\right). 
\label{glumass}
\end{equation}
The gluino mass
is allowed to be negative, as is the case in mAMSB, for example. Of course the
kinematic mass is just the absolute value, and the phase may be rotated away,
altering the phases of some of the Feynman rules. Negative masses for
neutralinos can also be rotated away in this way.

For the mixed neutralinos and charginos, the
external momentum is equal to the $\overline{DR}$ value of the diagonalised
mass.
All mixing angles and matrices are defined such that they diagonalise
the one-loop corrected mass matrix evaluated at the {\em minimum}
$\overline{DR}$ mass eigenvalue. In the case where inter-family flavour mixing
is not used in the sfermion sector, the above is also how the sfermion masses
and third family mixing is calculated. 
When flavour mixing between families is used, at
tree-level, the full flavour structure is present in sfermion masses.
Loop corrections are added only to the same family entries of the mass squared
matrix, not to
intra-family mixing entries. For the first two families, 
the external momentum is set equal to the $\overline{DR}$ sfermion mass
itself. For the third 
family, the external momentum is set equal to the the mass of the lightest
$\overline{DR}$ mass eigenvalue of the third family sfermion of the particular
flavour in question. 
All of the \SOFTSUSY~loop
corrections also neglect intra-family mixing themselves.

The pseudo-scalar Higgs  mass $m_{A^0}$ and CP-even Higgs masses $m_{h^0},
m_{H^0}$ are determined to full one-loop order
as in eq.~(E.6) of BMPZ in order to reduce
their scale dependence, which can be large~\cite{Katsikatsou:2000cd}. 
The zero-momentum $\oatq$, $\oabatau$, $\oabq$, $\oabas$, $\oatas$, $\oatauq$,
$\oatab$
2-loop corrections are 
also included in $m_{A^0}, m_{h^0},
m_{H^0}$~\cite{Degrassi:2001yf,Brignole:2001jy,Brignole:2002bz,Dedes:2003km}.
All one-loop corrections are included in
the determination of the charged Higgs pole mass.
Every Higgs mass is determined at an external momentum scale equal to its
$\overline{DR}$ mass. 

Finally, the running MSSM parameters are evolved back down to $M_Z$.
The whole process is iterated as shown in figure~\ref{fig:algorithm}, until the
$\overline{DR}$ sparticle masses evaluated at $M_S$
all converge to better than the desired fractional accuracy
(\code{TOLERANCE}), which may be set by the user in the main program or input
file. 

\subsection{Fine Tuning \label{sec:finetune}}

We now detail the fine-tuning calculation. 
As lower bounds on super partner masses are pushed up by colliders,
$m_{H_1}$ and $m_{H_2}$ may be forced to be much larger than $M_Z$ if they are
related to the other super particle masses, as is the case
for example in the case of 
minimal super gravity. 
If we re-phrase eq.~(\ref{mucond}) as
\begin{equation}
M_{\bar Z}^2  = -2
\mu^2 + \tan 2\beta \left[m_{\bar{H}_2}^2\tan \beta
- m_{\bar{H}_1}^2  \cot \beta \right],
\end{equation}
we see that the terms on the right-hand side must have 
some degree of cancellation in order to reproduce the 
observed value of $M_Z$.
But $\mu$ has a different origin to the SUSY breaking parameters and the
balancing appears unnatural. Various measures have
been proposed in order to quantify the apparent cancellation, for
example ref.s~\cite{Barbieri:1998uv,deCarlos:1993yy}.
The definition of naturalness $c_a$ 
of a `fundamental' parameter $a$ employed here is~\cite{deCarlos:1993yy} 
\begin{equation}
c_a \equiv \left| \frac{\partial \ln M_Z^2}{\partial \ln a} \right|.
\label{FT100index}
\end{equation}
From a choice of a set of fundamental parameters defined at the scale $M_X$:
$\{ a_i \}$, the 
fine-tuning of a particular model is defined to be $c=\mbox{max}(c_a)$.
$\{ a_i \}$ are any parameters in the user supplied boundary condition on the
soft supersymmetry breaking parameters augmented by $h_t(M_X), \mu(M_X)$ and
$m_3(M_X)$. The derivatives in eq.~(\ref{FT100index}) are calculated by 
numerically finding the derivative of $M_Z^{pole}=\hat M_Z +
\Re\Pi_{ZZ}^T(M_Z^2)$ in eq.~(\ref{mucond}). 
The input parameters are
changed slightly (one by one), then the MSSM parameter ensemble is run from 
$M_X$ to $M_S$ where the sparticle mass spectrum is determined along with the
corresponding $\overline{MS}$ Higgs VEV parameter
$v^2 \equiv v_1^2 + v_2^2$. 
First of all, $\tan \beta(M_S)$ is determined by inverting
eq.~(\ref{Bcond}) and the resulting value is utilised in a version of
eq.~(\ref{mucond}) inverted to give $M_Z^{pole}$ in terms of the other
parameters.
The resulting value of $M_Z^{pole}$ is the
prediction for the new changed input parameters, and its derivative is
determined by examining its behaviour as the initial changes in input
parameters tend to zero.

\appendix

\section{Running SOFTSUSY}
\label{sec:run}

A main program which produces an executable called
\code{softpoint.x}, is included in the
\SOFTSUSY~distribution. For the calculation of the spectrum of single points
in parameter space, we recommend the
SLHA~\cite{lhacc} input/output 
option. The user must modify a
file (e.g.\ \code{lesHouchesInput}, as provided in the standard distribution)
that specifies the input parameters. The user may then run the code with 
\small\begin{verbatim}
./softpoint.x leshouches < lesHouchesInput
\end{verbatim}\normalsize
In this case, the output will also be in
SLHA format. Such output can be used for input into other programs 
which subscribe to the accord, such as \code{SDECAY}~\cite{sdecay},
\code{PYTHIA}~\cite{pythia} (for 
simulating sparticle production and decays at colliders) or
\code{micrOMEGAs}~\cite{micromegas} (for
calculating the relic density of neutralinos, $b\rightarrow s \gamma$ and
$\mu \rightarrow e \gamma$), for example. For further details on the necessary
format of the input file, see ref.~\cite{lhacc} and appendix~\ref{sec:input}. 

If the user desires to quickly run a single parameter point in AMSB, mSUGRA or
GMSB parameter space, 
but does not wish to use the SLHA, the following options are available:
\small\begin{verbatim}
./softpoint.x sugra <m0> <m12> <a0> <tanb> <mgut> <sgnMu>
./softpoint.x amsb <m0> <m32> <tanb> <mgut> <sgnMu>
./softpoint.x gmsb <n5> <mMess> <lambda> <tanb> <sgnMu> 
\end{verbatim}\normalsize
Bracketed entries should be replaced by the desired numerical values, (in GeV
if they are dimensionful). The program will provide output from one point in
CMSSM, 
mAMSB or GMSB. 
If \code{<mgut>}~is specified as \code{unified}~on input, 
\SOFTSUSY~will determine
\code{mgut}~to 
be the scale that $g_1(M_{GUT})=g_2(M_{GUT})$, usually of order $10^{16}$ GeV.
If \code{<mgut>}~is set to be \code{msusy}, the SUSY breaking parameters will
be set at $M_{SUSY}$.

For users that are not familiar with \code{C++}, we note that the executable
interface allows the calculation at just one parameter point in SUSY breaking 
space. If scans are required, the user can either call \SOFTSUSY~from a
shell script or use a system call from a main \code{C}~program to the
executable. Alternatively, a main program showing an example of a scan is
provided. \code{C++}~beginners should note in
the following that ``method'' means function, that objects contain a list of
data structures and functions and that for a user to access (change
or reference) the data encoded in an object, one of its functions should be
called. Such functions are given in tables of the following appendices.

\subsection{Input file \label{sec:input}}

If, as recommended, the SLHA option is used for input, the user may
add
a \SOFTSUSY-specific block to the input file in the following format, with
bracketed entries replaced by double precision values:
\small
\begin{verbatim}
Block SOFTSUSY     # SOFTSUSY specific inputs
  1   <TOLERANCE>  # desired fractional accuracy in output
  2   <MIXING>     # quark mixing option
  3   <PRINTOUT>   # gives additional verbose output during calculation
  4   <QEWSB>      # change electroweak symmetry breaking scale
  5   <INCLUDE_2_LOOP_SCALAR_CORRECTIONS> # Full 2-loop running in RGEs
  6   <PRECISION>  # number of significant figures in SLHA output
  7   <numHiggsLoops> # number of loops in REWSB/mh calculation
 10   <forceSlha1>    # if =1, tries to force output into SLHA 1 format
\end{verbatim}
\normalsize
The fractional numerical precision on masses and couplings output by \SOFTSUSY~is better than \code{TOLERANCE}, which
sets the accuracy of the whole calculation. The iteration of
each physical SUSY particle mass is required to converge to a
fractional accuracy smaller than \code{TOLERANCE}. Sub-iterations are required
to converge to a better accuracy than $10^{-2}\times$\code{TOLERANCE} for
$s_W$ and 
$10^{-4}\times$\code{TOLERANCE} for $\mu$. The accuracy of the Runge-Kutta RGE
changes from iteration to iteration but is proportional to the value of 
\code{TOLERANCE}. Values between $10^{-2}$ and $10^{-6}$ are common, lower
values  
mean that \SOFTSUSY~takes significantly longer to perform the calculation.

The next parameter
\code{MIXING} determines what $M_Z$ boundary condition will be used for the
quark Yukawa matrix parameters.
\code{MIXING}=0.0 sets the quark mixings to zero but includes the first two 
family's diagonal terms.
\code{MIXING}=1.0,2.0 sets all the mixing at $M_Z$
to be in the up-quark or down-quark
sector respectively, as in eq.~(\ref{ckm}).

Setting \code{PRINTOUT} to a non-zero value gives additional information
on 
each successive iteration. If \code{PRINTOUT}$>$0, a warning flag is
produced when the overall iteration 
finishes. The predicted values of $M_Z^{pole}$ and $\tan \beta(M_S)$ after
iteration convergence
are also output\footnote{Note that the input value of $\tan \beta$ is the
value at $M_Z$.}. The level of convergence, $\mu(M_S)$, $m_3^2(M_S)$ and $M_Z$
are output with each iteration, as well as a flag if the object becomes
non-perturbative. 
\code{PRINTOUT}$>$1 produces output on the fine-tuning calculation. The
predicted values of $M_Z^{pole}$ and $\tan \beta(M_S)$ are output with each
variation 
in the initial inputs. A warning flag is produced when a
negative-mass squared scalar is present.
\code{PRINTOUT}$>$2 prints output on the sub-iterations that determine
$\mu(M_S)$ and $s_W(M_S)$, and flags the nature of any tachyons
encountered. Values \code{PRINTOUT}$>$0 are only required
if additional diagnostics are required for debugging purposes.

\code{QEWSB}~may be used to multiplicatively change the scale $M_S$ at which 
the Higgs potential is minimised and sparticle masses calculated, or it may
alternatively be used as a fixed scale.
This can be useful if one wants to examine
the scale dependence of the results~\cite{scaledep}. Setting
\code{QEWSB}=$x<M_Z/1$ GeV,
sets $M_S = x\sqrt{m_{\tilde t_1}(M_S) m_{\tilde t_2}(M_S)}$. Values from
0.5 to 2 are common. If $M_S<M_Z$ results from the above expression, $M_S=M_Z$
is used. If on the other hand a {\em fixed scale}\/ is desired from $M_S$, 
setting \code{QEWSB}$>M_Z/1$ GeV in the input results in 
$M_S=$\code{QEWSB}~GeV being fixed.

If \code{INCLUDE\_2\_LOOP\_SCALAR\_CORRECTIONS} is switched off (\code{0}) as
in the default case, 
2-loop RGEs~\cite{mandv}
 are used for the Higgs and gaugino masses, $\mu$, Yukawa and gauge
couplings but  
1-loop RGEs are used for other MSSM parameters. Switching on the 2-loop
corrections (\code{1})
results in a full 2-loop RGE evolution, but slows the calculation by a factor
of approximately three.
\code{<numHiggsLoops>} may be set to either \code{2} (default) or \code{1},
and is the number of loops of particles allowed to contribute to the Higgs
mass threshold calculation and the tadpoles in the electroweak symmetry
breaking conditions.

If \code{softpoint.x}~is used {\em without}\/ the SLHA interface, default
Standard Model inputs are used from the files \code{def.h} and \code{lowe.h}.
The low energy data is encoded in a \code{QedQcd} object and must be
provided. 
The default numbers supplied and contained in the \code{QedQcd} object 
are given in units of
GeV and running masses are in the $\overline{MS}$ scheme. 
For the bottom and
top masses, {\em either}\/ the running mass {\em 
  or}\/ the pole mass must be supplied as an input. The type of mass not given
for input is  calculated
by \SOFTSUSY~at the 3-loop QCD level. We recommend, along with
ref.~\cite{bottomMass}, that the
running mass be used for $m_b$ since there are smaller theoretical errors in
the 
extraction of this quantity from experiment than the pole mass. 
The scale dependent quantities in this object are then evolved to $M_Z$ by 
the method \code{toMz}, to provide the low-scale empirical boundary condition
for the rest of the calculation. The MSSM spectrum calculated depends most
crucially upon $M_Z, \alpha(M_Z), \alpha_s(M_Z)$ and input third family
fermion masses.

\section{Sample Program \label{sec:prog}}
We now present the sample program from which it is possible to run
\SOFTSUSY~in a simple fashion. The program we presents here performs a scan in
the variable $\tan \beta$, with other parameters as in point
CMSSM10.1.1~\cite{bench}. 
It then prints the four pole Higgs masses as a function of $\tan \beta$ in the
standard output channel. If there are any problems with the parameter point,
the program prints out these instead of the Higgs masses.
The most important features of the objects are 
described in appendix~\ref{sec:objects}.
The sample program \code{main.cpp}~has the following form: 
\small
\begin{verbatim}
#include <iostream>
#include "mycomplex.h"
#include "def.h"
#include "linalg.h"
#include "lowe.h"
#include "rge.h"
#include "softsusy.h"
#include "softpars.h"
#include "susy.h"
#include "utils.h"
#include "numerics.h"

int main() {
  /// Sets up exception handling
  signal(SIGFPE, FPE_ExceptionHandler); 

 /// Sets format of output: 6 decimal places
  outputCharacteristics(6);

  cerr << "SOFTSUSY" << SOFTSUSY_VERSION 
       << " test program, Ben Allanach 2002\n";
  cerr << "If you use SOFTSUSY, please refer to B.C. Allanach,\n";
  cerr << "Comput. Phys. Commun. 143 (2002) 305, hep-ph/0104145\n";

  /// Parameters used: CMSSM parameters
  double m12 = 500., a0 = 0., mGutGuess = 2.0e16, tanb = 10.0, m0 = 125.;
  int sgnMu = 1;      ///< sign of mu parameter 
  int numPoints = 10; ///< number of scan points

  QedQcd oneset;      ///< See "lowe.h" for default definitions parameters

  /// most important Standard Model inputs: you may change these and recompile
  double alphasMZ = 0.1187, mtop = 173.4, mbmb = 4.2;
  oneset.setAlpha(ALPHAS, alphasMZ);
  oneset.setPoleMt(mtop);
  oneset.setMass(mBottom, mbmb);

  oneset.toMz();      ///< Runs SM fermion masses to MZ

  /// Print out the SM data being used, as well as quark mixing assumption and
  /// the numerical accuracy of the solution
  cout << "# Low energy data in SOFTSUSY: MIXING=" << MIXING << " TOLERANCE=" 
       << TOLERANCE << endl << oneset << endl;

  /// Print out header line
  cout << "# tan beta   mh           mA           mH0          mH+-\n";

  int i; 
  /// Set limits of tan beta scan
  double startTanb = 3.0, endTanb = 50.0;
  /// Cycle through different points in the scan
  for (i = 0; i<=numPoints; i++) {

    tanb = (endTanb - startTanb) / double(numPoints) * double(i) +
      startTanb; // set tan beta ready for the scan.

    /// Preparation for calculation: set up object and input parameters
    MssmSoftsusy r; 
    DoubleVector pars(3); 
    pars(1) = m0; pars(2) = m12; pars(3) = a0;
    bool uni = true; // MGUT defined by g1(MGUT)=g2(MGUT)
    
    /// Calculate the spectrum
    r.lowOrg(sugraBcs, mGutGuess, pars, sgnMu, tanb, oneset, uni);

    /// check the point in question is problem free: if so print the output
    if (!r.displayProblem().test()) 
      cout << tanb << " " << r.displayPhys().mh0 << " " 
	   << r.displayPhys().mA0 << " " 
	   << r.displayPhys().mH0 << " " 
	   << r.displayPhys().mHpm << endl;
    else
      /// print out what the problem(s) is(are)
      cout << tanb << " " << r.displayProblem() << endl;
  }
}
\end{verbatim}
\normalsize
First of all, a function \code{signal}~is called which attempts to catch any floating point
exceptions that may occur during the running of the program. 
Then, after an initial introductory print-out, the variables specifying the
supersymmetry breaking parameters are specified. For these, the same notation
as appendix~\ref{sec:run} is used. Next, the important Standard Model inputs
are defined and combined with the defaults already present in the
\code{QedQcd} object. The top running mass is calculated from the pole mass
and 
Standard Model fermion masses and gauge couplings are
then run up to $M_Z$ with the method \code{toMz}.

\begin{table}\begin{center}
\begin{tabular}[l]{lll} \label{bcs}name & arguments \\ 
\hline
\code{sugraBcs} & $m_0$, $m_{1/2}$, $A_0$ \\
\code{amsbBcs} & $m_{3/2}$, $m_0$ \\
\code{gmsbBcs} & $n_5$, $m_{mess}$, $\Lambda$ \\
\end{tabular}\caption{SUSY breaking boundary conditions available to the user, detailing arguments
in order. The asterisk denotes additional information in the text.}\end{center}\end{table}

If \code{gaugeUnification=true}, \code{softsusy}~will determine
\code{mGutGuess} from electroweak gauge unification, using
the \code{mGutGuess} value supplied as an initial guess. 
The user can supply a void function that sets the supersymmetry breaking
parameters from an input \code{DoubleVector}. In the sample code given above,
this function is
\code{sugraBcs} and is applied to the \code{MssmSoftsusy} object at the
scale $M_{GUT}$, which will be determined by the user (but \code{mGutGuess}
will be
used as an initial guess). 
Other examples of available boundary conditions are given in
Table~\ref{bcs}. 
The user must supply a \code{DoubleVector} containing the
numerical values of the arguments, correctly ordered as in
Table~\ref{bcs}.
\code{sugraBcs}, for example, calls the \code{MssmSoftsusy} method
\code{standardSugra(m0, m12, a0)}, which sets all scalar masses equal to
\code{m0}, 
all gaugino masses to\footnote{With our sign conventions, the CMSSM should
  impose a common negative gaugino mass.} $-$\code{m12} and all trilinear
scalar couplings to 
\code{a0}, in the standard universal fashion. 
If the user desires to write his or her own boundary condition, it must
conform to the prototype\\
\code{void userDefinedBcs(MssmSoftsusy \& m, const DoubleVector \& inputs)}\\
The method \code{lowOrg} drives the calculation, after which $\tan \beta$ and
the Higgs masses are
printed out.

One other small extension to the SLHA input is the option of inputting
negative SUSY breaking scalar mass squared values as the theoretical boundary
condition at the input scale of the theoretical boundary condition. If this is
desired, one writes the {\em negative} of the mass in the SLHA input file.

\section{Sample Output \label{sec:output}}
For the recommended SLHA option, the conventions for the output are 
explained in Ref.~\cite{lhacc}. 
We present the non-SLHA compliant \SOFTSUSY~output 
for the main program above, which can be run by the command
\small\begin{verbatim} 
./softsusy.x
\end{verbatim}\normalsize
The output obtained was
\small\begin{verbatim}
# Low energy data in SOFTSUSY: MIXING=1 TOLERANCE=1.000000e-03
mU: 1.375003e-03  mC: 6.250222e-01  mt: 1.657605e+02  mt^pole: 1.734000e+02
mD: 2.734060e-03  mS: 5.986152e-02  mB: 2.866748e+00  mb(mb):  4.200000e+00
mE: 5.026664e-04  mM: 1.039355e-01  mT: 1.751578e+00  mb^pole: 4.962029e+00
aE: 1.279250e+02  aS: 1.187000e-01   Q: 9.118760e+01  mT^pole: 1.776990e+00
loops: 3        thresholds: 1

# tan beta   mh           mA           mH0          mH+-
3.000000e+00 1.039255e+02 8.453169e+02 8.473537e+02 8.494758e+02
7.700000e+00 1.140726e+02 7.321935e+02 7.326559e+02 7.368674e+02
1.240000e+01 1.153050e+02 7.101467e+02 7.103521e+02 7.149075e+02
1.710000e+01 1.156316e+02 6.913273e+02 6.914275e+02 6.961964e+02
2.180000e+01 1.157377e+02 6.695505e+02 6.696477e+02 6.745775e+02
2.650000e+01 1.157523e+02 6.443757e+02 6.444185e+02 6.496148e+02
3.120000e+01 1.157003e+02 6.148784e+02 6.149922e+02 6.204020e+02
3.590000e+01 1.155505e+02 5.820500e+02 5.820491e+02 5.879275e+02
4.060000e+01 1.151542e+02 5.447311e+02 5.448647e+02 5.510808e+02
4.530000e+01 [ stau tachyon ]
5.000000e+01 [ stau tachyon ]
\end{verbatim}\normalsize

Firstly, the output details the input parameters, starting with
\code{MIXING}~and \code{TOLERANCE}.  
After the output of the input \code{QedQcd} object,
various pole Higgs masses are displayed for different values of tan beta, as
labelled by the various columns. The columns are, respectively, 
$\tan \beta$, $m_{h^0}$, $m_{A^0}$, $m_{H^0}$ and $m_{H^\pm}$.
In the case that there is a problem with the point, the higgs masses are
replaced by a warning: three points have tachyonic staus, and the last has
not achieved convergence of the iteration and possesses a negative mass
squared for $m_{H^\pm}$. Here, \code{stau tachyon}~indicates that the
stau mass has become imaginary and so the scalar potential minimum does not 
conserve electromagnetism. For the highest value of $\tan \beta$, \code{No
  convergence}~indicates that \SOFTSUSY~was not able to iterate the
calculation to the desired accuracy. \code{hpm tachyon}~indicates that the
charged Higgs masses have become imaginary and the scalar potential minimum
does not conserve electromagnetism. The presence of such tachyons rules such
parameter points out.  
We list and explain all of the possible problem flags in 
the following section.

\subsection{Problem flags \label{sec:prob}}
Any associated problems such as negative mass-squared scalars
or 
inconsistent EWSB are flagged at the end of the output.
We now list the problems, indicating their meaning:
\begin{itemize}
\item
If \code{No convergence} appears, then \SOFTSUSY~is indicating that 
it didn't achieve the accuracy of \code{TOLERANCE} within less than 40
iterations. The output of the code is therefore to be considered unreliable
and it is not clear from the output whether the point is allowed or
disallowed, despite the presence or absence of other warning messages. 
This error flag often appears near the boundary of electroweak symmetry
breaking, (where $\mu(M_{SUSY})=0)$), where the iterative algorithm is not
stable. To 
calculate the position of the electroweak symmetry boundary, one should
interpolate between regions a small distance away from it. 
\item
\code{Non-perturbative}
indicates that \SOFTSUSY~encountered couplings reaching 
Landau poles when 
evolving, and could not calculate any further. Any results obtained using
perturbation theory (for example those of \code{SOFTSUSY})
therefore cannot be trusted.
\item
\code{Infra-red quasi fixed point breached} indicates that the parameter point
is at a Landau  pole of a Yukawa coupling. This should not be a problem
provided no other errors are flagged.
\item
\code{muSqWrongSign} indicates that the Higgs minimisation conditions imply
that $\mu^2<0$, meaning that the desired electroweak minimum is not present in
the model. The model is ruled out.
\item
\code{m3sq} indicates that $m_3^2$ from eq.~(\ref{Bcond}) has the incorrect
sign, meaning that the desired electroweak minimum is not present in
the model. The model is ruled out.
\item
The \code{tachyon} variable labels if a scalar particle other than the Higgs
has acquired a negative mass squared, when $M_Z^2<0$ or when 
a pole Higgs masses is imaginary. The model is ruled out.
\code{tachyon} is of an enumerated type \code{tachyonType}, which relabels the
integers $z \in (0,\ 15)$ to indicate which particle is a tachyon according to
the rule
\begin{eqnarray}
0&=&\mbox{no tachyon},\ 1=\tilde e,\ 2=\tilde \mu,\ 3=\tilde \tau,\ 
4=\tilde u,\ 5=\tilde c,\ 6=\tilde c,\ 7=\tilde d,\ 
8=\tilde s, \nonumber \\
9&=&{\tilde b},\ 10=h^0,\ 11=A^0,\ 12=H^\pm,\ 13={\tilde \nu}_e,\
14={\tilde \nu}_\mu,\ 15={\tilde \nu}_\tau,\nonumber
\end{eqnarray}
respectively.
\item
\code{noRhoConvergence}~is flagged when \SOFTSUSY~cannot calculate the $\rho$
parameter and determine gauge couplings from data, typically because of
tachyons or infinities that have crept into the calculation. The other
problems are serious enough to rule the model out.
\item
\code{higgsUfb}~and \code{bProblem}~indicate that the desired electroweak minimum is in fact a
saddle point of the potential, thus the model is ruled out.
\item
\code{mgutOutOfBounds}~is flagged if the value of the gauge unification scale
predicted by eq.~\ref{mguteq}
is outside the range $10^4$ GeV $< M_X < 5 \times 10^{17}$ GeV. The GUT-scale
has been set to the appropriate limit, and the \SOFTSUSY~numbers cannot be
trusted. 
\item
\code{inaccurateHiggsMass}~is flagged when the $\overline{DR}$ perturbation
series has broken down, and higher order terms are potentially $\sim
\mathcal{O}(1)$. This can happen when $(M_3/m_{{\tilde t}_1})^2>16 \pi^2$, or
when 
$(\mu/m_{{\tilde t}_1})^2>16 \pi^2$~\cite{Degrassi:2001yf}. The rest of the calculation should be fine,
but one cannot trust the higgs masses. One should use some other program that
uses the on-shell scheme to calculate Higgs masses for these
points\footnote{Although we note that there are many different points for
  which the 
$\overline{OS}$ scheme has a perturbation series problem and one must use
the  $\overline{DR}$ instead.}.
\end{itemize}
Thus flags other than \code{No convergence}, \code{Infra-red quasi fixed point
  breached}, {\code inaccurateHiggsMass} or
\code{Non-perturbative} indicate an
unphysical minimum of the scalar potential, effectively ruling the model point 
out.
Another structure within \code{MssmSoftsusy} of type \code{sProblem} flags
various potential problems 
with the object, for example the lack of radiative EWSB
or negative mass squared scalars (excluding the Higgs mass squared
parameters). 
This structure is shown in table~\ref{table:probs}. In addition, the method
\code{test} 
prints out if any of the possible data variables flagging problems are true.
The \code{higgsUfb} flag is true if
\begin{equation}
m_{H_1}^2 + 2 \mu^2 + m_{H_2}^2 - 2 |m_3^2| <0 \label{higgsufb}
\end{equation}
is not satisfied, implying that the desired electroweak minimum is either a
maximum or a saddle-point of the tree-level Higgs
potential~\cite{Barger:1994gh}. The contents of \code{sPhysical} and
\code{sProblem} can be output with overloaded \code{<<} operators.
\code{noConvergence} means that the desired
accuracy was not reached. \code{nonPerturbative} or \code{irqfp} flags the
existence of a Landau pole in the renormalisation group evolution, and the
calculation is not perturbatively reliable so any results should be discarded.
All other problems except \code{noConvergence} and
\code{nonPerturbative} should be
considered as grounds for ruling the model out.
\code{noRhoConvergence} occurs when the pseudo-scalar Higgs $A^0$ has 
a negative mass squared (i.e.\ an invalid electroweak vacuum).

\section{Switches and Constants \label{sec:switches}}
The file \code{def.h} contains the switches and constants. If they are
changed, the code must be 
recompiled in order to use the new values. \code{def.cpp} contains initial
values for global variables (in the \SOFTSUSY~namespace). These may be changed
by the user in their main programs. 
Table~\ref{tab:switches} shows the
most important parameters in 
\code{def.h}, detailing the default values that the constants have.
$G_\mu$ and $M_Z$ have been obtained using the latest particle
data group numbers~\cite{Groom:2000in}. \vspace{1cm}
\begin{table}\begin{center}
\begin{tabular}{lll} \label{tab:switches}
variable & default & description \\ \hline
\code{ARRAY\_BOUNDS\_CHECKING} & off & Vector and Matrix bounds checking\\
\code{EPSTOL} & $10^{-11}$ & Underflow accuracy\\
\code{GMU} & 1.16637 10$^{-5}$ & $G_\mu$, Fermi constant from muon decay\\
\code{MZCENT} & 91.1876 & Pole mass of the $Z^0$ boson $M_Z$.\\
\end{tabular}\caption{Switches and constants. Starred entries have more explanation in the text.
$G_\mu$ is in units of GeV$^2$ and $M_Z$ in GeV.} \end{center}\end{table}

\section{Object Structure\label{sec:objects}}

We now go on to sketch the objects and their relationship. This is necessary
information for generalisation beyond the MSSM\@.
Only methods and data which are deemed important for prospective users are
mentioned here, but there are many others within the code itself.

\subsection{Linear Algebra}

The \SOFTSUSY~program comes with its own linear algebra classes: 
\code{Complex}, \code{DoubleVector}, \code{DoubleMatrix},
\code{ComplexVector}, \code{ComplexMatrix}. 
Constructors of the latter four objects involve the dimensions of the object,
which start 
at 1. \code{Complex} objects are constructed with their real and imaginary
parts respectively.
For example, to define a vector $a_{i=1,2,3}$, a matrix $m_{i=1\ldots3,
j=1\ldots 4}$ of type \code{double} and a Complex number $b=1-i$:
\small
\begin{verbatim}
  DoubleVector a(3);
  DoubleMatrix m(3, 4);
  Complex      b(1.0, -1.0); 
\end{verbatim}
\normalsize
Obvious algebraic operators between these classes (such as
multiplication, addition, subtraction) are defined with overloaded
operators \code{*}, \code{+}, \code{-} respectively. Elements of the vector
and matrix classes are referred to with 
brackets \code{()}. \code{DoubleVector} and \code{DoubleMatrix} classes are
contained within each of the higher level objects that we now describe. 

\subsection{General Structure}

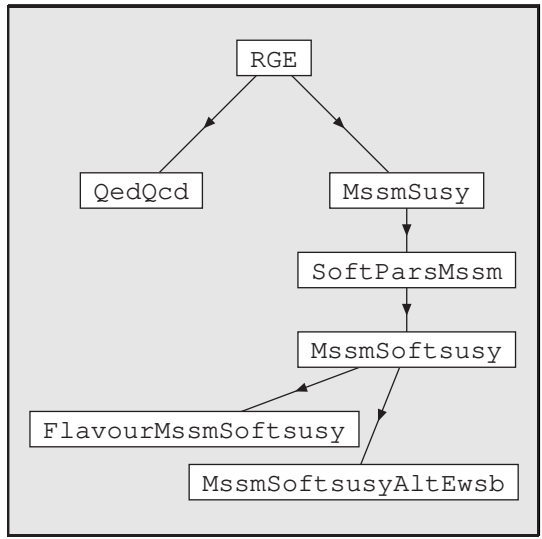
\begin{figure}\begin{center} \label{fig:objstruc}
\begin{picture}(200,200)
\GBox(200,200)(0,0){0.9}
\ArrowLine(100,180)(50,130)
\ArrowLine(100,180)(150,130)
\ArrowLine(150,130)(150,100)
\ArrowLine(150,100)(150,70)
\ArrowLine(150,70)(130,20)
\ArrowLine(150,70)(70,40)
\SetPFont{Teletype}{10}
\put(0,0){\framebox(200,200){}}
\BText(100,180){RGE}
\BText(50,130){QedQcd}
\BText(150,130){MssmSusy}
\BText(150,100){SoftParsMssm}
\BText(150,70){MssmSoftsusy}
\BText(130,20){MssmSoftsusyAltEwsb}
\BText(70,40){FlavourMssmSoftsusy}
\end{picture}
\caption{Heuristic high-level object structure of \SOFTSUSY\@. Inheritance is
displayed by the lines.}\end{center}\end{figure}
From a RGE point of view, a particular
quantum 
field theory model consists of a set of couplings and masses defined at some
renormalisation scale $\mu$. A set of $\beta$ functions describes the evolution
of the parameters and masses to a different scale $\mu'$. This concept
is embodied in an {\em abstract} \code{RGE} object, which contains the
methods required to run objects of derived classes to different
renormalisation scales. The other objects displayed in
figure~\ref{fig:objstruc} are particular instances of \code{RGE}, and
therefore inherit from it.
\code{QedQcd} objects consist of data on the 
quark and lepton masses and gauge couplings.
They contain the $\beta$ functions for
running in an effective QED$\times$QCD theory below $m_t$.
An object of class \code{MssmSusy} contains the
Yukawa couplings, and the three gauge couplings
of the MSSM\@. It also contains the superpotential $\mu$ term (not
to be confused with the renormalisation scale), $\tan \beta$, the ratio of
the two Higgs doublet VEVs as well as $v=\sqrt{v_1^2+v_2^2}$. Its $\beta$ functions are
valid in the exact SUSY limit of the MSSM\@. 
The major part of the code resides within the \code{MssmSoftsusy} class.
Objects of this type have
all the functionality of \code{MssmSusy}, with soft SUSY breaking terms 
and theoretical boundary conditions contained in the inherited
class~\code{SoftParsMssm}. 
It also contains an object of type \code{QedQcd} which contains weak scale
empirical data.
Code in the \code{MssmSoftsusy} class organises and performs the main part of
the calculation.
\code{MssmSoftsusyAltEwsb} objects are a slight variant of {\tt MssmSoftsusy}:
it takes non-universal Higgs mass boundary conditions at the SUSY breaking
scale. 
\code{FlavourMssmSoftsusy} objects are also slight variants: they have
full flavour-mixed output and input.
In the following, we provide basic information on classes so that users
may program using \SOFTSUSY\@. Highly detailed and technical
documentation on the program may be obtained from the \SOFTSUSY~website.

\subsection{\code{RGE}~Class}

The data and important methods in \code{RGE} are presented in
table~\ref{tab:rge}. 
Each of the higher level objects described in this appendix have explicitly
named \code{display} 
and \code{set} methods that are used to access or change the data contained
within each object. In table~\ref{tab:rge} (as in the following tables in
this section), these accessing methods are listed on the same row as the
relevant data variable. 

The \code{RGE} method \code{runto(mup, eps)} will automatically run any
derived object to the scale \code{mup} with a fractional accuracy of evolution
\code{eps}. In order to define this evolution, any object that inherits from an
{\tt RGE} must contain three methods: 
\code{display},
\code{set}, \code{beta} shown in table~\ref{tab:rge}. 
\code{DoubleVector display() const}
must return a vector containing all masses and couplings of the object, in some
arbitrary user-defined order. 
\code{void set(const DoubleVector \& v)} must set
these couplings given a 
\code{DoubleVector v} defined in the same order as the display function.
\code{DoubleVector beta() const} must then return the $\beta$ functions in a
\code{DoubleVector}  
defined as
\begin{equation}
\beta_i = \frac{d a_i}{d \ln \mu},
\end{equation}
where $a_i$ denotes any mass or coupling of the model. The ordering of the
$a_i$ must be identical in each of the three methods.

\begin{table}\begin{center}
\begin{tabular}{lll} \label{tab:rge}
data variable & & methods \\ \hline
\code{double mu}$=\mu$ & renormalisation scale & \code{setMu} \\
& (GeV) &  \code{displayMu} \\ \hline
\code{int numpars} & number of scale dependent & \code{setPars} \\
 & parameters & \code{howMany} \\ \hline
\code{int loops} & accuracy of RGE & \code{setLoops} \\
 & & \code{displayLoops} \\ \hline
\code{int thresholds} & accuracy level of threshold &
\code{setThresholds} \\  & computation & \code{displayThresholds}\\
 & & \\
method & \multicolumn{2}{l}{function} \\ \hline
\code{DoubleVector display()} & \multicolumn{2}{l}{displays all running
parameters (*)} \\ 
\code{void set(DoubleVector)} & \multicolumn{2}{l}{sets all running
parameters (*)} \\ 
\code{DoubleVector beta} & \multicolumn{2}{l}{displays beta functions of all
running parameters (*)} \\ 
\code{runto} & \multicolumn{2}{l}{runs object to new value of
\code{mu}} \\ 
\end{tabular}\caption{Abstract \code{RGE} class. (*) indicates that derived objects {\em must}\/
contain these methods (see text). }\end{center}\end{table}

\subsection{\code{QedQcd}~Class}

\begin{table}\begin{center}\begin{tabular}{lll} \label{tab:qedqcd}
data variable & & methods \\ \hline
\code{DoubleVector a} & $\overline{MS}$ gauge couplings & \code{setAlpha} \\
$\alpha(\mu),\alpha_s(\mu)$& &  \code{displayAlpha} \\ \hline
\code{DoubleVector m} & running fermion masses & \code{setMass}
\\ 
$m_f(\mu)$& vector (1\ldots 9) (GeV)&  \code{displayMass} \\ \hline
\code{double mtPole, mbPole} & pole top/bottom/tau & \code{setPoleMt},
\code{setPoleMb} \\
\code{double mtauPole} & mass & \code{setPoleMtau} \\
$m_t^{pole}, m_b^{pole}, m_\tau^{pole}$&(GeV) &  \code{displayPoleMt} \code{displayPoleMb}
\\
 & & \code{displayPoleMtau} \\
 &  & \\
method & \multicolumn{2}{l}{function} \\ \hline
\code{runGauge} & \multicolumn{2}{l}{runs gauge couplings {\em only}}\\
\code{toMt, toMZ} & \multicolumn{2}{l}{runs fermion masses and gauge couplings
} \\
 & \multicolumn{2}{l}{from $Q'$ to $m_t^{pole}$ or $M_Z$}\\
\end{tabular}\caption{\code{QedQcd} class. $Q'$ is defined in the text.}\end{center}\end{table}
The \code{QedQcd} class contains a \code{DoubleVector} of quark and lepton 
$\overline{MS}$ masses
($m_f=m_{u,d,e,c,s,\mu,t,b,\tau}(\mu)$), as shown in table~\ref{tab:qedqcd}.
Its contents may be printed to standard output or read from standard input
(with the same format in each 
case) by using the operators
\code{<<} or \code{>>}, as can all the non-abstract objects mentioned in this
section. The methods \code{toMz()}, \code{toMt()} act on an
initial object defined with each fermion mass $m_f$ defined at a scale
\begin{equation}
Q' = \mbox{max} (1 \gev, m_f(m_f)) \label{scaley}
\end{equation}
and gauge couplings at $M_Z$. 

\subsection{\code{MssmSusy}~Class \label{sec:mssmsusy}}
\begin{table}\begin{center}\begin{tabular}{lll} \label{tab:mssmsusy}
data variable & & methods \\ \hline
\code{DoubleMatrix u, d, e}  & Yukawa
couplings & \code{setYukawaElement} \\ 
$(Y_U)_{ij}, (Y_D)_{ij}, (Y_E)_{ij}$&(3 by 3 matrix) &
\code{setYukawaMatrix} \\
 & & \code{displayYukawaElement}\\
 & & \code{displayYukawaMatrix}\\ \hline
\code{DoubleVector g} & MSSM gauge couplings & \code{setAllGauge} \\
$g_i$ & ($1 \ldots 3$) vector & \code{setGaugeCoupling} \\ 
 & & \code{displayGauge} \\
 & & \code{displayGaugeCoupling} \\ \hline
\code{double smu} & bilinear Higgs superpotential & \code{setSusyMu}\\
$\mu$ & parameter & \code{displaySusyMu}\\ \hline
\code{double tanb} & ratio of Higgs VEVs (at  & \code{setTanb}\\
  $\tan \beta$  &  current renormalisation scale)& \code{displayTanb} \\ \hline
\code{double hVev} & Higgs VEV & \code{setHvev} \\
$v$ & & \code{displayHvev} \\
 & & \\
method & \multicolumn{2}{l}{function} \\ \hline 
\code{setDiagYukawas} & \multicolumn{2}{l}{calculates and sets all diagonal
Yukawa couplings }\\
& \multicolumn{2}{l}{given fermion masses and a Higgs VEV} \\
\code{getMasses} & \multicolumn{2}{l}{calculates quark and lepton masses from
Yukawa} \\
 &  \multicolumn{2}{l}{couplings}\\
\code{getQuarkMixing} & \multicolumn{2}{l}{mixes quark Yukawa couplings from mass to weak basis}\\
\code{getQuarkMixedYukawas} & \multicolumn{2}{l}{sets all entries of quark Yukawa couplings given
fermion} \\
 & \multicolumn{2}{l}{masses, Higgs VEV and CKM matrix }\\
\end{tabular}
\caption{\code{MssmSusy} class. }\end{center}\end{table}
The operators \code{<<}, \code{>>} have been overloaded to write or read a
\code{MssmSusy} object to/from a file stream. Table~\ref{tab:mssmsusy} shows
the data variables and important methods contained in the class. For the
Yukawa and gauge couplings, methods exist to either set (or display) one
element or a whole matrix or vector of them.

\subsection{\code{SoftParsMssm}~Class\label{sec:softie}}
The operators \code{<<}, \code{>>} have been overloaded to write or read a
\code{softParsMssm} object to/from a file stream. 
Table~\ref{tab:softparsmssm} shows
the data variables and important methods contained in the class.
\code{addAmsb()} adds anomaly mediated supersymmetry breaking
terms~\cite{Randall:1998uk} to 
the model's soft parameters. Such terms are proportional to the VEV of a
compensator superfield, so $m_{3/2}$ in table~\ref{tab:softparsmssm} must have
been set before \code{addAmsb} is used. 
\code{minimalGmsb(int n5, double lambda, double mMess)} applies the messenger
scale \code{mMess} boundary conditions to the soft masses in minimal
gauge-mediated supersymmetry breaking~\cite{gmsb}. \code{n5} denotes the
number of $5\oplus \bar 5$ messenger fields that are present and
\code{lambda}($\Lambda$) is as described in ref.~\cite{gmsb}. 

\begin{table}\begin{center}\begin{tabular}{lll} \label{tab:softparsmssm}
data & & methods \\ \hline
\code{double m32} &compensator VEV$^*$ & \code{setM32} \\
$m_{3/2}$ &(GeV) & \code{displayGravitino}\\ \hline
\code{DoubleVector mGaugino} & (1 \ldots 3) vector of gaugino &
\code{setGauginoMass} \\
$M_{1,2,3}$ & mass parameters & \code{displayGaugino} \\ \hline
\code{DoubleMatrix ua,da,ea} & (3 by 3) matrix of trilinear&
\code{setTrilinearElement}\\
$U_A,D_A,E_A$ & soft terms (GeV) &  \code{displayTrilinearElement}\\
 & & \code{displaySoftA} \\ \hline
\code{DoubleMatrix mQLsq} & (3 by 3) matrices of soft &
\code{setSoftMassElement} \\
\code{mURsq,mDRsq,mLLsq} & SUSY breaking masses & \code{setSoftMassMatrix}
\\
\code{mSEsq} & (GeV$^2$) & \code{displaySoftMassSquared} \\ 
$(m_{\tilde Q_L}^2)$, $(m_{\tilde u_R}^2)$, $(m_{\tilde d_R}^2)$,
& & \\ 
$(m_{\tilde L_L}^2)$, $(m_{\tilde e_R}^2)$  & & \\ \hline
\code{double m3sq,mH1sq,mH2sq} & Bilinear Higgs parameters & \code{setM3Squared}\\
$m_3^2$, $m_{H_1}^2$, $m_{H_2}^2$ & (GeV, GeV$^2$, GeV$^2$) & \code{setMh1Squared}\\
 & & \code{setMh2Squared} \\
 & & \code{displayM3Squared} \\
 & & \code{displayMh1Squared} \\
 & & \code{displayMh2Squared} \\ 
\\
method & \multicolumn{2}{l}{function}\\ \hline
\code{standardSugra} & \multicolumn{2}{l}{Sets all universal soft terms}\\
\code{universalScalars} & \multicolumn{2}{l}{Sets universal scalar masses}\\ 
\code{universalGauginos} & \multicolumn{2}{l}{Sets universal gaugino masses}\\ 
\code{universalTrilinears} & \multicolumn{2}{l}{Sets universal soft breaking trilinear
couplings}\\  
\code{addAmsb} & \multicolumn{2}{l}{Adds AMSB soft terms to current
object$^*$}\\ 
\code{minimalGmsb} & \multicolumn{2}{l}{Gauge-mediated soft terms used as
boundary conditions$^*$}\\ 
\hline
\end{tabular}
\caption{\code{SoftParsMssm} class data and methods. The asterisk denotes additional information in the text.}\end{center}\end{table}

\subsection{\code{MssmSoftsusy}~Class \label{sec:mssmsoftsusy}}

\begin{table}\begin{center}\begin{tabular}{ll} \label{tab:sphys}
data variable & description \\ \hline
\code{double mh0,mA0,mH0,mHpm} & $h^0, A^0, H^0,
H^\pm$masses \\
\code{DoubleVector msnu} & vector of $m_{{\tilde \nu}_{i=1 \ldots 3}}$ masses \\
\code{DoubleVector mch,mneut} & vectors of $m_{{\chi^\pm}_{i=1 \ldots 2}}$, 
$m_{{\chi^0}_{i=1 \ldots 4}}$ respectively \\
\code{double mGluino} & gluino mass $m_{\tilde g}$ \\
\code{DoubleMatrix mixNeut} & 4 by 4 orthogonal neutralino mixing matrix $O$
\\
\code{double thetaL, thetaR} & $\theta_{L, R}$ chargino mixing angles \\
\code{double thetat, thetab} & $\theta_{t,b}$ sparticle mixing angles \\
\code{double thetatau, thetaH} & $\theta_{\tau}, \alpha$ sparticle and Higgs
mixing angles \\
\code{DoubleMatrix mu, md, me} & (2 by 3) matrices of up squark, down squark
and\\
 &  charged slepton masses \\
\code{double t1OV1Ms, t2OV2Ms} & 2-loop tadpoles $t_1/v_1$ and $t_2 / v_2$
evaluated at $M_S$ \\
\code{double t1OV1Ms1loop, t2OV2Ms1loop} & 1-loop tadpoles $t_1/v_1$ and $t_2 / v_2$
evaluated at $M_S$ \\
\end{tabular}
\caption{\code{sPhysical} structure. Masses are pole masses, and stored in units of
GeV. Mixing angles are in radian units.}\end{center}\end{table}

\code{MssmSoftSusy} objects contain a structure \code{sPhysical} encapsulating
the physical information on the superparticles (pole masses and physical
mixings), as shown in table~\ref{tab:sphys}.  
Another structure of type \code{drBarPars} inherits from \code{sPhysical} but
instead contains information on $\overline{DR}$ masses and mixing angles. 
\code{MssmSoftSusy} objects
also contain one of these structures for calculational convenience:
the information is used in order
to calculate loop corrections to various masses.
\begin{table}\begin{center}\begin{tabular}{ll} \label{tab:drbarpars}
data variable & description \\ \hline
\code{double mt, mb, mtau} & Third family fermion masses \\
$m_t(Q), m_b(Q), m_\tau(Q)$ &                     \\ \hline
\code{DoubleVector mnBpmz, mchBpmz} & Absolute neutralino and chargino masses\\
$m_{\chi_i^0}, m_{\chi^\pm_i}$ & (1\ldots4, 1\ldots2) vectors \\ \hline
\code{ComplexMatrix nBpmz} & Neutralino mixing matrix \\
$N$          & (4 by 4 complex matrix) \\ \hline
\code{ComplexMatrix uBpmz, vBpmz} & Chargino mixing matrices \\
$U, V$       & (2 by 2 complex matrices) \\ \hline
 & \\
name         & function \\ \hline
\code{mpzNeutralino} & Gives mixing matrices required to make \\ & neutralino
masses positive$^*$\\ 
\code{mpzChargino} & Gives mixing matrices required to make \\ & chargino
masses positive$^*$\\ 
\end{tabular}
\caption{\code{drBarPars} structure. Masses are in the $\overline{DR}$ scheme, and
  stored in units of GeV. Mixing angles are in radian units. Functions marked
  with an asterisk are mentioned in the text.}\end{center}\end{table}
As table~\ref{tab:drbarpars} shows, a
method \code{mpzCharginos} returns the 2 by 2 complex diagonalisation
matrices $U,V$ that result in positive $\overline{DR}$ chargino masses, as
defined in ref.~\cite{Pierce:1997zz}. 
The method
\code{mpzNeutralinos} is present in order to convert $O$ to the complex matrix
$N$ defined in ref.~\cite{Pierce:1997zz} that would produce only positive
$\overline{DR}$ neutralino masses. This information, as well as
$\overline{DR}$ third family
fermion masses are stored in the \code{drBarPars} structure.

\begin{table}\begin{center}\begin{tabular}{ll} \label{table:probs}
data variable & flags \\ \hline
\code{mgutOutOfBounds} & $M_X>5 \times 10^{17}$ GeV or $M_X<10^4$ GeV\\
\code{irqfp} & in a region with a Landau pole \\
\code{noRhoConvergence} & the $\rho$ iterative routine doesn't converge\\
\code{noConvergence} & the main iteration routine doesn't converge \\ 
\code{tachyon} & a non-Higgs scalar has negative mass squared \\
\code{muSqWrongSign} & $\mu^2$ from eq.~(\ref{mucond}) negative \\
\code{m3sq} & $m_3^2$ from eq.~(\ref{Bcond}) has incorrect sign \\
\code{higgsUfb} & eq.~(\ref{higgsufb}) is not satisfied\\
\code{nonperturbative} & a Landau pole was reached below the unification scale
\\
\code{noMuConvergence} & $\mu$ could not be calculated reliably\\
\code{inaccurateHiggsMass} & Higgs masses cannot be trusted\\
\code{problemThrown} & numerical exception occurred during run (infinities etc)\\
\end{tabular}
\caption{\code{sProblem} structure. All data variables are boolean values except for
  \code{tachyon}, which is of type \code{tachyonType}. See
  Section~\protect\ref{sec:prob} for more details.}\end{center}\end{table}

\begin{table}\begin{center}\begin{tabular}{lll} \label{tab:mssmsoftdata}
data & & methods \\ \hline
\code{double mwPred} & pole $M_W$ prediction & \code{setMw} \\
$M_W$ & (GeV) & \code{displayMw} \\ \hline
\code{double msusy} & Minimisation scale & \code{displayMsusy} \\
$M_{S}$ & (GeV) & \code{setMsusy} \\ \hline
\code{QedQcd dataset} & $M_Z$ boundary condition on  &
\code{setData} \\
 & Standard Model couplings & \code{displayDataSet} \\ \hline
\code{sProblem problem} & problem flags & \code{displayProblem} \\
&& \code{flagIrqfp}, \code{flagB} \\
&& \code{flagNonperturbative} \\
&& \code{flagTachyon}, \code{flagHiggsufb} \\
&& \code{flagNoConvergence} \\
&& \code{flagNoMuConvergence} \\
&& \code{flagNoRhoConvergence} \\
&& \code{flagMusqwrongsign} \\
&& \code{flagAllProblems} \\ \hline
\code{DrBarPars forLoops} & $\overline{DR}$ masses and mixings& \code{displayDrBarPars} \\
 &  & \code{setDrBarPars} \\ \hline
\code{sPhysical physpars} & pole masses and mixings & \code{displayPhys}\\
 && \code{setPhys}\\
\end{tabular}\caption{\code{MssmSoftsusy} class data and accessor methods.}\end{center}\end{table}

\begin{table}\begin{center}\begin{tabular}{ll} \label{tab:softmeth}
name & function \\ \hline
\code{lowOrg} & Driver routine for whole calculation$^*$\\ 
\code{methodBoundaryCondition} & Boundary condition for derived objects$^*$\\
\code{itLowsoft} & Performs the iteration between $M_Z$ and \\
 &unification scale\\  
\code{sparticleThresholdCorrections} & $\overline{DR}$ radiative corrections
to Standard Model \\
 & couplings at $M_Z$\\ 
\code{physical} & Calculates sparticle pole masses and mixings\\ 
\code{calcDrBarPars} & Calculates $\overline{DR}$ pole masses and mixings\\ 
\code{rewsb} & Sets $\mu$, $B$ from EWSB conditions\\ 
\code{fineTune} & Calculates fine-tuning for soft parameters$^*$ \\ & and
$h_t$\\ 
\code{getVev} & Calculates VEV $v^{\overline{DR}}$ at current scale from $Z$\\
 & self-energy and gauge couplings\\ 
\code{calcSinthdrbar} & Calculates $s_W^{\overline{DR}}$ at current scale
from\\  & gauge couplings \\
\code{calcMs} & Calculates $M_S$\\ 
\code{printShort} & short list of important parameters printed out \\ &to standard
output in columns\\ 
\code{printLong} & long list of important parameters printed out \\ &to standard
output in columns\\ 
\code{outputFcncs} & prints a list of flavour-changing $\delta$
parameters$^*$\\
\code{lesHouchesAccordOutput} & prints output in SLHA~\cite{lhacc} format\\
\code{sinSqThetaEff} & calculates then returns $\sin^2\theta_{eff}^l$\\
\end{tabular}
\caption{\code{MssmSoftsusy} methods and related functions. Functions marked with an
asterisk are mentioned in the text.}\end{center}\end{table}

\code{MssmSoftsusy} data variables and accessors can be viewed in
table~\ref{tab:mssmsoftdata} and the most important high-level 
methods are displayed in table~\ref{tab:softmeth}.
The operators \code{<<}, \code{>>} have been overloaded to write or read 
\code{MssmSoftusy} objects or \code{sPhysical} structures to/from a file
stream. 
The driver routine for the RGE evolution and unification calculation is
\small
\begin{verbatim}
double MssmSoftsusy::lowOrg
(void (*boundaryCondition)(MssmSoftsusy &, const DoubleVector &),
 double mxGuess, const DoubleVector & pars, int sgnMu, double tanb, const QedQcd &
 oneset, bool gaugeUnification, bool ewsbBCscale = false)
\end{verbatim}
\normalsize
The user-supplied \code{boundaryCondition} function sets the soft parameters 
according to 
the elements of the supplied \code{DoubleVector} at \code{mxGuess}, 
as discussed in appendix~\ref{sec:prog}. If \code{gaugeUnification}~is
\code{true}, the scale that unifies the electroweak gauge couplings is used
and returned by the function. If \code{gaugeUnification}~is \code{false}, the
function simply returns \code{mxGuess}.
\code{ewsbBCscale}~is an optional argument: if one wishes to impose the soft
SUSY breaking boundary conditions at $M_{SUSY}$ rather than $M_X$, one should
call \code{lowOrg}~with the value \code{true}~for this argument. Omitting the
argument, or giving it the default \code{false}~value means that $M_X$ will be
used instead. 
\code{pars} contains a \code{DoubleVector}
of soft SUSY breaking parameters to be applied as the theoretical boundary
condition.  
\code{sgnMu} is the sign of the
superpotential $\mu$ parameter, \code{tanb} is the value of $\tan \beta (M_Z)$
required and \code{oneset} contains the $M_Z$ scale low energy data.

The fine tuning (as defined in sec.~\ref{sec:calculation}) can be
calculated with the method 
\small
\begin{verbatim}
DoubleVector MssmSoftsusy::fineTune(void (*boundaryCondition)
        (MssmSoftsusy &, const DoubleVector &), const DoubleVector 
         & bcPars, double mx) const 
\end{verbatim}
\normalsize
This function should only be applied to an \code{MssmSoftsusy} object which
has been processed by \code{lowOrg}. \code{mx} is the unification scale and
\code{boundaryCondition} is the function that sets the unification scale soft
parameters, as discussed above. 
In derived objects, the virtual method
\code{methodBoundaryCondition} may be used to set data additional to
\code{MssmSoftsusy} from the \code{boundaryCondition} function. 
The method outputs the fine-tuning of a
parameter $a_{i=1\ldots n}$ in the \code{bcPars(n+3)} \code{DoubleVector}, 
with the $(n+1,n+2,n+3)^{th}$ element
of \code{bcPars} being the fine-tuning with respect to the Higgs potential
parameters ($\mu$ and $B$) and
the top Yukawa coupling
$(h_t)$ respectively. \code{fineTune} is an optional feature.
\code{sinSqThetaEff()} returns a \code{double} number corresponding to a full
one-loop calculation of the quantity $\sin^2 \theta_{eff}^l$. It does {\em
  not}\/ contain any 2-loop corrections, and may not be accurate enough for
precision electroweak fits.

\subsection{\code{MssmSoftsusyAltEwsb}~Class \label{sec:altewsb}}
The \code{MssmSoftsusyAltEwsb}~class, which inherits directly from
\code{MssmSoftsusy}, adds two private data variables: \code{muCond}~and
\code{mAcond}, both massive parameters in units of GeV. They fix the
boundary conditions on $\mu(M_{SUSY})$ and the pole pseudo-scalar Higgs
mass $m_A(\mbox{pole})$. In this case, $m_A(M_{SUSY})$ is extracted 
from the input $m_A(\mbox{pole})$ via $m_A^2(M_{SUSY}) = m_A^2(\mbox{pole})
 + \Pi_{AA}(M_{SUSY})$, where $\Pi_{AA}(M_{SUSY})$ is the MSSM pseudo-scalar 
self-energy correction.
The Higgs mass squared parameters
$m_{H_{1,2}}^2$ are {\em not}\/ set at $M_{GUT}$: they are instead set at
$M_{SUSY}$ by solving the simultaneous equations~\ref{mucond},~\ref{Bcond}
and using the relationship between $m_3^2(M_{SUSY})$ and $m_A(M_{SUSY})$:
\begin{eqnarray}
m_{H_1}^2 &=& \sin^2 \beta (m_A^2 + M_Z^2) + \frac{t_1}{v_1} (1 - \sin^4
\beta)
- \sin^2 \beta \cos^2 \beta \frac{t_2}{v_2} - (\mu^2 + \frac{1}{2} M_Z^2) 
\nonumber \\
m_{H_2}^2 &=& \cos^2 \beta (m_A^2 + M_Z^2) + \frac{t_2}{v_2} (1 - \cos^4
\beta)
- \sin^2 \beta \cos^2 \beta \frac{t_1}{v_1} - (\mu^2 + \frac{1}{2} M_Z^2),
\label{altEwsbEqs}
\end{eqnarray}
where all quantities in eq.~\ref{altEwsbEqs} are running parameters evaluated
at $M_{SUSY}$. 
This option is covered under the SLHA input parameters
\code{EXTPAR} 
\code{23,26}~\cite{lhacc}.

\subsection{\code{FlavourMssmSoftsusy}~Class \label{sec:flavour}}
\begin{table}\begin{center}\begin{tabular}{lll} \label{tab:flavsoft}
data & & methods \\ \hline
\code{flavourPhysical fv}&sfermion mixing/mass data & \code{setFlavourPhys} \\
                        &                           &\code{displayFlavourPhys}\\ 
\hline
\code{double theta12, theta23} & CKM matrix parameters &
\code{setTheta12} \\
\code{double theta13, deltaCkm} & & \code{setTheta13} \\
$\theta_{12}, \theta_{23}, \theta_{13}, \delta$ & & \code{setTheta23} \\
 & & \code{setDelta} \\
 & & \code{displayTheta12} \\
 & & \code{displayTheta23} \\
 & & \code{displayTheta13} \\
 & & \code{displayDelta} \\
\hline
\code{double thetaB12, thetaB23} & PMNS matrix parameters &
\code{setThetaB12} \\
\code{double thetaB13} & & \code{setThetaB13} \\
${\bar \theta}_{12},{\bar \theta}_{13},{\bar \theta}_{23}$ & &
\code{setThetaB23} \\ 
 & & \code{displayThetaB12} \\
 & & \code{displayThetaB23} \\
 & & \code{displayThetaB13} \\
\hline
\code{double mNuE, mNuMu, mNuTau} & light neutrino masses& \code{setMnuE}\\
$m_{\nu_e}, m_{\nu_\mu}, m_{\nu_\tau}$ & & \code{setMnuNu} \\
 & & \code{setMnuTau} \\
 & & \code{displayMnuE} \\
 & & \code{displayMnuMu} \\
 & & \code{displayMnuTau} \\ \hline
\code{double mcMc} & running charm quark mass & \code{setMcMc} \\
$m_c(m_c)$ & & \code{displayMcMc} \\
\code{double md2GeV, mu2GeV, ms2GeV} & quark masses at 2 GeV &
\code{setMd2GeV}, \code{setMu2GeV} \\
$m_d(2), m_u(2), m_s(2)$ 
 & & \code{setMs2GeV} \\
 & & \code{displayMd2GeV} \\
 & & \code{displayMu2GeV} \\
 & & \code{displayMs2GeV} \\ \hline
\code{double mePole, mmuPole} & pole lepton masses & \code{setPoleMe,
  setPoleMmu} \\
$m_e^{pole}, m_\mu^{pole}$ & & \code{displayPoleMe} \\
 & & \code{displayPoleMmu} \\
\end{tabular}\caption{\code{FlavourMssmSoftsusy} class data and accessor methods. All angles are
  measured in radians and masses are measured in GeV.}\end{center}\end{table}
\code{FlavourMssmSoftsusy}~objects inherit directly from
\code{MssmSoftsusy}~with the addition of the variables shown in
Table~\ref{tab:flavsoft}.
Some of these
constitute the angles of the PMNS and CKM matrices in the standard
parameterisation of 
eq.~\ref{ckmPar}. The
\code{lesHouchesAccordOutput} method has been overloaded to provide flavour
violating input and output in accordance with the SLHA2
conventions~\cite{lhacc2}. Note that \SOFTSUSY~currently does {\em not}\/
contain CP-violating complex phases, despite the inclusion of $\delta$. Thus,
when CKM angles are input via the 
SLHA2 in the Wolfenstein parameterisation, the magnitude of the 13 entry 
is fit to $\sin \theta_{13}$ as in eq.~\ref{ckm} and $\delta$ is set to zero
(for now). The relevant method is
\small
\begin{verbatim}
void FlavourMssmSoftsusy::setAngles
(double lambda, double aCkm, double rhobar, double etabar) 
\end{verbatim}
\normalsize
Currently, an identical parameterisation is used for the PMNS
matrix that describes lepton mixing, except $\theta_{ij}$ is replaced by
${\bar \theta}_{ij}$. There is currently no provision for a CP-violating phase. 
Effective light neutrino masses 
are also included in the object. 

An optional feature intended for studies of flavour-changing neutral
currents (FCNCs) is the method 
\begin{verbatim}
void FlavourMssmSoftsusy::sCkm
(DoubleMatrix & deltaULL, DoubleMatrix & deltaURR, DoubleMatrix & deltaULR, 
 DoubleMatrix & deltaDLL, DoubleMatrix & deltaDRR, DoubleMatrix & deltaDLR) 
const
\end{verbatim}
which calculates the parameters $(\delta^{u,d}_{LL,LR,RR})_{ij}$ calculated in
the mass-squared-insertion 
approximation (after a rotation to the super CKM basis), as defined
in Ref.~\cite{Gabbiani:1996hi}. 

The relationship between the super CKM basis of the mass matrices (including
one-loop corrections) and the pole-mass basis as described by SLHA2, is
contained in the structure 
\code{flavourPhysical}. We list the relevant data in Table~\ref{tab:flavphys}.
\begin{table}\begin{center}\begin{tabular}{lll} \label{tab:flavphys}
data & description \\ \hline
\code{DoubleMatrix dSqMix} & 6$\times$6 down squark mixing \\
\code{DoubleMatrix uSqMix} & 6$\times$6 up squark mixing \\
\code{DoubleVector msD} & 6 mass-ordered down squark masses\\
\code{DoubleVector msU} & 6 mass-ordered up squark masses\\
\code{DoubleMatrix eSqMix} & 6$\times$6 charged slepton mixing matrix \\
\code{DoubleMatrix nuSqMix} & 3$\times$3 sneutrino mixing matrix \\
\code{DoubleVector msE} & 6 mass-ordered charged slepton masses  \\
\code{DoubleVector msNu} & 3 mass-ordered sneutrino masses \\
\end{tabular}\caption{\code{flavourPhysical} structure. All mass parameters
  are pole 
  masses, stored in
  units of GeV. The mixing matrix definitions exactly coincide with
  those in the 
SLHA2~\protect\cite{lhacc2} and describe the transformation between the
loop-corrected super CKM basis and the mass basis.}\end{center}\end{table}
Note that, in the case that CMSSM type inputs are listed in the SLHA2 {\em
  as well as} the flavour input parameters $\hat T_{U,D,E}$ (assumed to be in
the superCKM basis), a special procedure is required to implement both types
of term.
At $M_X$, the $A_0$ terms are added in the interaction basis. There is then a
transformation to the super CKM basis, where any values of $\hat T_{U,D,E}$
input over-write the trilinear terms. Finally, the trilinear terms are
transformed back to the interaction basis.

\acknowledgments
This work has been partially supported by PPARC, STFC and the Aspen Center for
Physics.
We would like to thank
A.~Djouadi, S.~Kraml and W. Porod 
for help with detailed comparisons~\cite{comparison} with the codes
{\tt SUSPECT}, {\tt ISAJET} and {\tt SPHENO},
S.~Akula, P.~Athron, M.~Baak, M.~Badziak, 
G.~B\'elanger, F.~Borzumati, F.~Boudjema, F.~Brummer,
M.~Eads,  
M.~Hamer,
J.~Hetherington, M.~Ibe, J.~Kersten, S.~Kom, S.~Krippendorf, S.~Kulkarni,
K.~Ishikawa, 
R.~Lu, S.~Martin, P.~Meade,
N.~Mahmoudi,  
A.~Pukhov, M.~Ramage,
R.~Ruiz, 
S.~Sekmen, A.~Sheplyakov, P.~Slavich, P.~Skands, J.~Smidt, C.~Tamarit, 
S.~Tetsuo, 
T.~Watari and A.~Wingerter
for bug finding and coding suggestions, I.~Gafton, J. Holt, F. Krauss,
D. Sanford and F. Yu for
suggestions on the draft, D.~Ross for  
information on the Passarino-Veltman integrals and K.~Matchev for 
other useful discussions. 

\providecommand{\href}[2]{#2}\begingroup\raggedright\endgroup

\end{document}